\documentclass{article}

\usepackage{arxiv}

\usepackage[utf8]{inputenc} 
\usepackage[T1]{fontenc}    
\usepackage{hyperref}       
\usepackage{url}            
\usepackage{booktabs}       
\usepackage{amsfonts}       
\usepackage{nicefrac}       
\usepackage{microtype}      
\usepackage{lipsum}		
\usepackage{graphicx}
\usepackage{natbib}
\usepackage{doi}
\usepackage{framed,multirow}
\usepackage{graphicx}
\usepackage{xcolor}
\usepackage[multi-part-units=single]{siunitx}
\usepackage{amsmath}
\usepackage{amssymb}
\usepackage{bm}
\usepackage{url}
\usepackage{arydshln}
\usepackage{booktabs}
\usepackage{amssymb}
\usepackage{latexsym}
\usepackage{ulem}
\usepackage{float}
\usepackage{placeins}
\usepackage{url}
\usepackage{xcolor}
\usepackage{hyperref}
\usepackage{pifont}
\newcommand{\cmark}{\ding{51}}%
\newcommand{\xmark}{\ding{55}}

\title{Guided Reconstruction with Conditioned Diffusion Models for Unsupervised Anomaly Detection in Brain MRIs}

\author{ Finn Behrendt\\
	Hamburg University of Technology\\
	Hamburg, Germany 
	\And
    Debayan Bhattacharya \\
	Hamburg University of Technology\\
	Hamburg, Germany 
  	\And
    Robin Mieling \\
	Hamburg University of Technology\\
	Hamburg, Germany 
 	\And
    Lennart Maack  \\
	Hamburg University of Technology\\
	Hamburg, Germany 
 	\And
    Julia Krüger \\
	Jung Diagnostics GmbH \\
	Hamburg, Germany 
 	\And
    Roland Opfer \\
	Jung Diagnostics GmbH \\
	Hamburg, Germany 
 	\And
    Alexander Schlaefer \\
	Hamburg University of Technology\\
	Hamburg, Germany }

\date{}

\renewcommand{\headeright}{Preprint}
\renewcommand{\undertitle}{}
\renewcommand{\shorttitle}{Behrendt et al. \textit{Conditioned Diffusion Models for UAD in Brain MRIs}}

\begin{document}
\maketitle

\begin{abstract}
The application of supervised models to clinical screening tasks is challenging due to the need for annotated data for each considered pathology. Unsupervised Anomaly Detection (UAD) is an alternative approach that aims to identify any anomaly as an outlier from a healthy training distribution. A prevalent strategy for UAD in brain MRI involves using generative models to learn the reconstruction of healthy brain anatomy for a given input image. As these models should fail to reconstruct unhealthy structures, the reconstruction errors indicate anomalies. However, a significant challenge is to balance the accurate reconstruction of healthy anatomy and the undesired replication of abnormal structures. While diffusion models have shown promising results with detailed and accurate reconstructions, they face challenges in preserving intensity characteristics, resulting in false positives.
We propose conditioning the denoising process of diffusion models with additional information derived from a latent representation of the input image. We demonstrate that this conditioning allows for accurate and local adaptation to the general input intensity distribution while avoiding the replication of unhealthy structures. We compare the novel approach to different state-of-the-art methods and for different data sets. Our results show substantial improvements in the segmentation performance, with the Dice score improved by 11.9\%, 20.0\%, and 44.6\%, for the BraTS, ATLAS and MSLUB data sets, respectively, while maintaining competitive performance on the WMH data set. Furthermore, our results indicate effective domain adaptation across different MRI acquisitions and simulated contrasts, an important attribute for general anomaly detection methods. The code for our work is available at \url{https://github.com/FinnBehrendt/Conditioned-Diffusion-Models-UAD}

\end{abstract}


\section{Introduction}
Magnetic Resonance Imaging (MRI) is an important tool for diagnosing various conditions in the human brain \cite{vernooij2007incidental,LUNDERVOLD2019102}. However, interpreting brain MRIs can be error-prone, time-consuming, and places a significant workload on available radiologists \cite{bruno2015understanding,mcdonald2015effects}. To address these challenges and improve diagnostic efficiency, deep learning techniques like convolutional neural networks (CNN) have shown great promise in assisting radiologists by automating certain aspects of the analysis \cite{LUNDERVOLD2019102}. A common task is the detection and delineation of pathological structures in the MRI scans such as tumors \cite{perkuhn2018clinical}, white matter lesions \cite{moeskops2018evaluation} or Alzheimer's disease \cite{islam2018brain}. Supervised deep learning approaches exhibit robust performance for these tasks, given that task-specific, annotated data sets are available. However, gathering such data sets is a cumbersome and costly process. Additionally, applying supervised models for screening tasks is difficult as any pathology must be detected, even those that are underrepresented or not included in the training data. In this context, screening tasks refer to scenarios where various conditions need to be identified without a specific target condition in mind. This includes assisting radiologists in detecting a wide range of findings, from expected abnormalities to unexpected or incidental ones, as well as applications in large population studies, such as the Hamburg City Health Study \cite{Jagodzinski.2020}.\\
An alternative approach is unsupervised anomaly detection (UAD), which relies on healthy data instead of annotated pathologies. The goal is to learn the underlying data distribution of healthy brain MRI scans and to identify anomalies as outliers from that learned distribution.  A popular approach is reconstruction-based UAD, where generative models (GM) are trained to reconstruct healthy anatomy. Given that the GMs are trained exclusively on healthy data, they should fail to generate unhealthy components of the input images and replace them by approximations of healthy anatomy. Subsequently, the discrepancy of input and pseudo-healthy reconstruction is measured, e.g., by the mean absolute error where large errors indicate anomalies \cite{baur2021autoencoders,kascenas2021denoising,chen2020unsupervised,pinaya2022unsupervised}. Therefore, in theory, any deviation from the learned norm can be localized, including known conditions, unexpected anomalies such as artifacts, or previously unseen pathologies.\\
However, in practice, reconstruction-based UAD methods typically do not result in perfect anomaly detection and segmentation. Considering that the training task is reconstructing the input image, a key challenge is to limit the  reconstruction to healthy brain anatomy. On the one hand, GMs that generate highly accurate reconstructions tend to perform a ’copy task’, resulting in unhealthy structures still reflected in the pseudo-healthy reconstructions. These unhealthy structures do not deviate from the input image, resulting in false negatives in the segmentation mask.
On the other hand, GMs with limited reconstruction accuracy may produce imperfect reconstructions everywhere, which in turn appear as differences in the anomaly map, even for healthy structures. This complicates distinguishing between actual pathologies and reconstruction errors, typically causing false positives in the segmentation map. Another related challenge in UAD is the potential domain shift between the distribution of the healthy data and the unhealthy test cases. Here, the reconstruction may appear different due to the domain shift; e.g., in simple cases, it is generally brighter than the input, and this shift reflects differences in the anomaly map.
In summary, the challenge is to avoid copying unhealthy input features to the reconstruction while still adapting the general appearance of the reconstruction to the input.\\
Recently, denoising diffusion probabilistic  models (DDPMs)  \cite{ho2020denoising} have shown promise as GMs for reconstruction-based UAD in brain MRI \cite{pinaya2022fast,wyatt2022anoddpm,BehrendtpDDPM}. DDPMs generate images by denoising images corrupted by artificial noise, leveraging a high-dimensional latent space to preserve spatial context and achieve high-fidelity reconstructions. 
While the spatial latent space enables the accurate reconstruction of healthy structures, the additional denoising task aims to prevent the DDPMs from solely copying the content of the input image. Therefore, DDPMs can achieve a reasonable trade-off between the reconstruction accuracy of healthy and unhealthy structures \cite{wyatt2022anoddpm,BehrendtpDDPM}.
However, a significant challenge remains in accurately reconstructing healthy brain anatomy that exhibits aligned intensity characteristics with the input image. The forward and backward processes of DDPMs do not adequately capture the highly variable local intensity distributions of MRI scans. This can result in discrepancies between the input and the reconstruction, which are difficult to distinguish from those arising from actual pathologies. This leads to reduced segmentation performance, particularly when facing domain shifts at test time. One potential solution could be to incorporate the input image as an additional input to the denoising process in the DDPM, e.g., as a second input channel. However, this could allow the denoising process to replicate the content of the input image, which would contradict the principles of reconstruction-based UAD. 
\\
In response to these challenges, we propose the use of conditioned DDPMs (cDDPMs) for UAD in brain MRI. Our approach includes a conditioning mechanism that guides the denoising process of the DDPM by utilizing an extra feature representation of the input image. This feature representation, derived from a CNN-based image encoder, does not capture detailed structural information. However, it contains the coarse local intensity information from the input image, which is often partially lost during the DDPM’s forward process. This way, we aim to align the local intensity distribution of the reconstructed image without providing detailed structural information that could be used to replicate unhealthy structures.
We conduct a comprehensive investigation of our conditioning approach, specifically addressing the challenges associated with reconstruction-based UAD in brain MRI. Initially, we evaluate the quality of healthy brain MRI reconstructions and assess the ability to reconstruct healthy anatomy while replacing unhealthy structures. Subsequently, we examine the domain adaptation capabilities of our method by assessing the alignment of intensity between input and reconstruction for datasets not encountered during training and by simulating varying contrast levels. Finally, we evaluate the segmentation performance of our approach on a variety of datasets, comparing established state-of-the-art UAD methods.
\\
Our results indicate that our conditioning approach effectively aligns the local intensity distributions of input and reconstruction without supporting the replication of unhealthy structures. Furthermore, cDDPMs can effectively adapt to different intensity and contrast profiles of different MRI data sets. 
As a result, our proposed cDDPMs address key challenges of DDPMs in reconstruction-based UAD in brain MRI and improve or match the segmentation performance of the compared baseline models. When compared to DDPMs, the Dice score significantly increases (p<0.05), rising from 50.27 to 56.30 for the BraTS21 dataset and from 20.18 to 24.22 for the ATLAS v2 dataset. For the MSLUB dataset, the performance improves from 9.71 to 14.04 and for the WMH dataset, both models report similar performance, with Dice scores of 12.06 and 11.59, respectively.\\

In summary, the main contributions of this work are:
\begin{itemize}
    \item Reconstruction Quality: We develop a conditioning mechanism within cDDPMs that guides the denoising process of DDPMs while avoiding the replication of unhealthy structures.
    \item Domain Adaptation and Intensity Alignment: Our conditioning mechanism aligns the local intensity distribution of the reconstructed image with that of the input image. This effectively improves the generalization to intensity shifts of different MRI scans.
    \item Segmentation Performance: The accurate reconstructions and aligned local intensity distributions featured by our conditioning mechanism can significantly improve the segmentation performance and applicability of DDPMs for UAD in brain MRI. 
\end{itemize}

This paper is organized as follows: In Section \ref{sec:recentwork}, we review relevant literature on UAD in brain MRI. In Section \ref{sec:methods}, we introduce DDPMs and subsequently explain our conditioning approach. In Section \ref{sec:expsetup}, we provide details of the experimental setup. In Section \ref{sec:results}, we present the results and subsequently discuss them in Section \ref{sec:discussion}. Finally, we provide a conclusion in Section \ref{sec:conclusion}.

\begin{figure*}[h!]
    \centering
    \includegraphics[width=\linewidth]{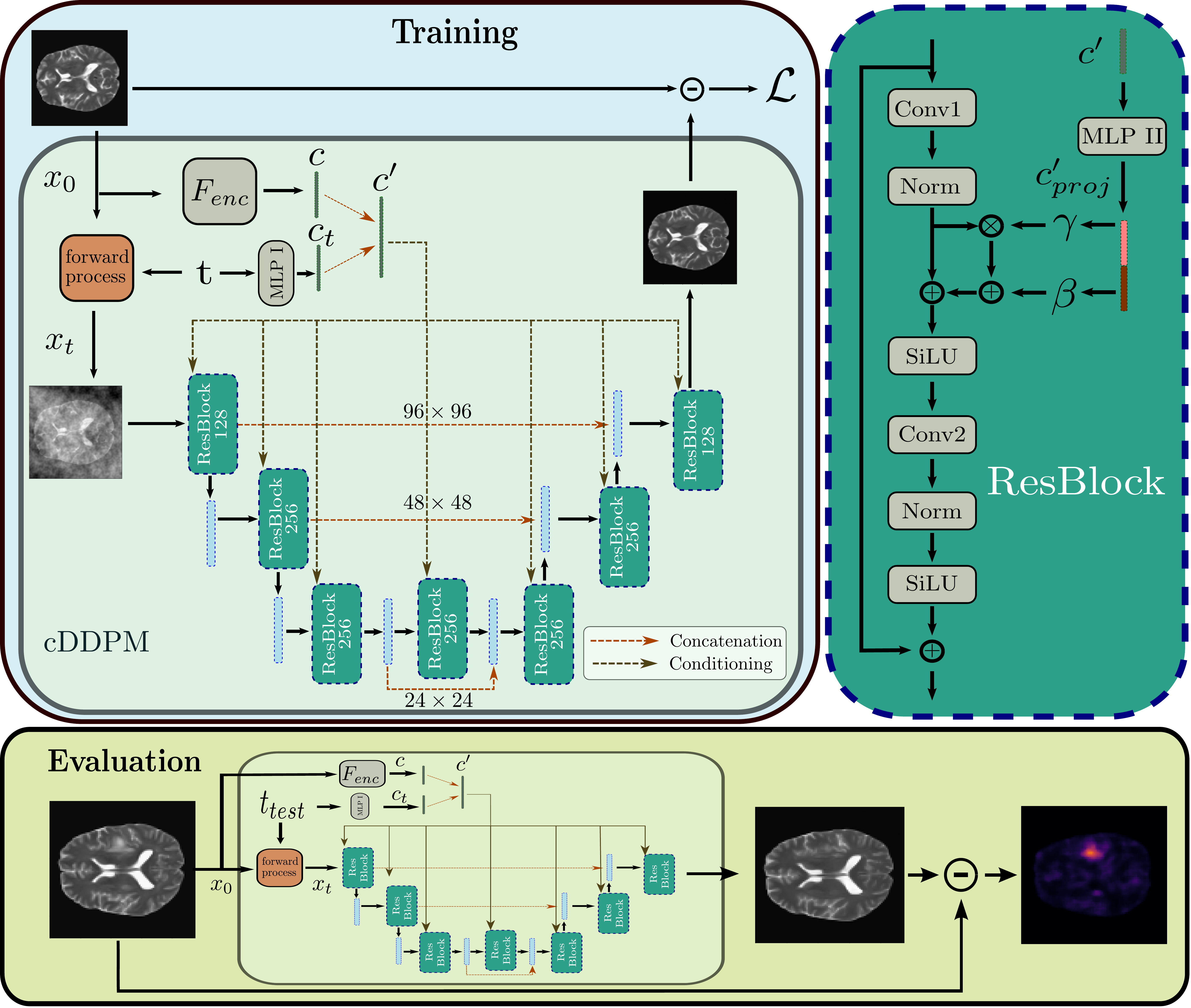}
    \caption{Overview of our proposed approach. The encoder representations are learned along with the DDPM in the main training stage to condition the denoising process. The timestep embedding $c_t$ is concatenated with the input image's projected encoder representation $c$. The resulting conditioning vector $c'$ is used to scale and shift feature maps of the denoising Unet in the residual blocks. Each residual block consists of two convolution operators (conv1, conv2), group normalization (Norm) and Sigmoid Linear Units (SiLU). During evaluation, the residual map between unhealthy brain images and their healthy reconstructions is used for anomaly detection.  }
    \label{fig:graphical abstract}
\end{figure*}

\section{Recent Work}   \label{sec:recentwork}
Autoencoders (AE) have been the primary focus of research on reconstruction-based UAD in brain MRI. Although these models exhibit potential in capturing the underlying healthy distribution, their effectiveness in UAD is limited by their blurry reconstructions \cite{baur2021autoencoders}.
To overcome this limitation, researchers have focused on improving the representations and reconstructions by adding skip connections with dropout \cite{baur2020bayesian}, using multi-scale features \cite{baur2020scale}, utilizing feature activation maps \cite{SILVARODRIGUEZ2022102526} or employing feature discrepancies \cite{Meissen.2022FAE,Deng.2022}. Additionally, online outlier removal strategies \cite{Behrendt_impured} have been proposed for AEs. In parallel, Variational Autoencoders (VAE) have been investigated for the UAD task \cite{Zimmerer.2019.VAE}, focusing on enhancing the used context in 2D \cite{zimmerer2018context} and 3D \cite{bengs2021three,behrendt2022capturing} or utilizing restoration methods \cite{chen2020unsupervised}. Moreover, Soft-Intro VAES (SIVAE) \cite{bercea2023generalizing} have been investigated. Additionally, Generative Adversarial Networks (GAN) have been explored for the task of UAD either as pure GAN \cite{han2021madgan,Schlegl.2019} or in combination with VAEs \cite{baur2018deep,Bercea.2023MICa} and VQ-VAEs with transformers \cite{pinaya2022unsupervised}. \\
While AEs with skip connections and a spatial latent space enable reconstructions of high fidelity, they tend to perform a 'copy task' which enables the reconstruction of unhealthy anatomy and therefore contradicts the UAD principle \cite{baur2021autoencoders, bercea2023generalizing}. Lately, \cite{kascenas2021denoising} have shown that AEs with skip connections can be effectively used for UAD in brain MRI if they are regularized by an additional denoising task. Taking a similar direction, DDPMs have shown promise in the field of UAD \cite{wyatt2022anoddpm,pinaya2022fast,graham2022denoising,bercea2023mask}. DDPMs provide high reconstruction fidelity, but important information about the input image can be lost due to the noising process. To address this, \cite{BehrendtpDDPM} proposed patch-based DDPMs that allow the use of parts of the original image content to provide information for the reconstruction of the input image. However, this patching strategy increases complexity and computational effort and can lead to artifacts in overlapping patch regions. A more efficient approach is seen in conditioning the denoising process of DDPMs with knowledge of the input image. Conditioned DDPMs have been successful in text-to-image synthesis tasks \cite{dhariwal2021diffusion} and image-guided synthetic image generation \cite{saharia2022palette,wang2022pretraining}. However, in the specific case of UAD, the objective is not to generate new images or to transfer styles but to accurately estimate a given input image while ensuring that unhealthy anatomy is absent in the estimation. Directly conditioning DDPMs with information from the input image can pose a risk of reconstructing unhealthy anatomy. Recent studies leverage different approaches to fuse information from a target image into the generation process of DDPMs.
\cite{wolleb2022diffusion} introduce a classifier model trained to differentiate between healthy and tumorous brain MRI scans. 
At its core, this model employs classifier guidance during the sampling process, targeting the denoising towards a 'healthy' classification. Similarly, \cite{sanchez2022healthy} apply classifier conditioning for counterfactual generation to remove tumors by conditioning the model to generate 'healthy' images. Despite showing promise, these methods rely on sample-level label information of given pathologies, contradicting the unsupervised setting. Moreover, their reliance on a scalar variable for conditioning limits the capacity to capture complex image characteristics.
The authors \cite{zhang2023iccv} approach this challenge by incorporating synthetic anomalies in two separate diffusion processes. Their method involves feeding concatenated noised images (normal and anomalously altered) into a denoising Unet. This approach relies on synthetic defects that can be added to the input images. While effective in industrial defect detection, as demonstrated on the MVTEC dataset, this approach's dependency on synthetic anomalies raises concerns about its applicability to medical datasets. 
\cite{mousakhan2023anomaly} propose an approach to fuse information of a target image into the generation process. By estimating and applying the same noise pattern to both the input and target images, their model aims to minimize the variance between these noised versions. While guiding the generation towards a target image, this method results in the loss of input image information, as the same noise is applied to the target image.\\
Our approach, detailed in the following sections, addresses these limitations by introducing a simple conditioning mechanism applied to the denoising Unet within DDPMs.  In contrast to \cite{wolleb2022diffusion} and \cite{sanchez2022healthy}, we utilize spatial conditioning information and rely on unlabelled data. Unlike the proposed conditioning in \cite{zhang2023iccv}, our approach does not rely on simulated anomalies and we directly condition the denoising process without the need for additional sub-networks. This direct conditioning of the denoising process in the Unet diverges from \cite{mousakhan2023anomaly}'s approach, ensuring that valuable information from the target image is retained and utilized effectively throughout the diffusion process. 
\section{Methods} \label{sec:methods}
In our proposed cDDPM we use an image encoder network and embed the input image in a context vector $\bm{c} \in \mathbb{R}^d$ to condition the denoising Unet on meaningful features of the input image. 
Our motivation is that the additional information in $\bm{c}$ guides the generation process towards consistent intensity characteristics across the input image and its reconstruction. Hence, by introducing the context vector $\bm{c}$ we aim to recover local intensity information lost during the forward (noising) process of DDPMs. We utilize an image encoder with a dense latent space to extract information regarding the coarse shape and local intensity information of the noise-free input image. This latent representation can then be used to condition the denoising process and supplement the individual context of the input image without providing detailed pixel-wise information that could be used to perform a 'copy task'. A general depiction of our approach is shown in Figure \ref{fig:graphical abstract}.

\subsection{DDPMs}
DDPMs are generative models that learn the underlying data distribution of images $\bm{x} \in \mathbb{R}^{H,W,C}$ with height $H$, width $W$ and $C$ channels, given a training set. Training of DDPMs consists of two steps. The forward process, where an input image $\bm{x}_0$ is gradually transformed to Gaussian noise $\bm{x}_T = \bm{\epsilon} \sim \mathcal{N}(\textbf{0},\textbf{I})$ and the backward process, where reversing the forward process is learned. \\
In the forward process, transforming $\bm{x}_0$ to $\bm{x}_T$ follows a predefined schedule $\beta_1,...,\beta_T$, where intermediate versions $\bm{x}_t$ are derived as
\begin{align*}
\bm{x}_t \sim q(\bm{x}_t|\bm{x}_0)=\mathcal{N}(\sqrt{\bar\alpha_t} \bm{x}_0,(1-\bar\alpha_t) \textbf{I}),  \\ \text{with\,\,} \, \bar\alpha_t=\prod\nolimits_{s=0}^{t}(1-\beta_t).
\end{align*}
The time step $t$ controls the amount of added noise and is sampled from $t \sim Uniform(1,...,T)$. For edge cases, the image $\bm{x}_t$ is transformed to pure noise ($t=T$) or no transformation is applied ($t=0$).
In the backward process, the reconstructed image $\bm{x}_0^{rec}$ is recovered from $\bm{x}_t$ by
\begin{align*}
\bm{x}_{0}^{rec} \sim p(\bm{x}_T)\prod\nolimits^T_{t=1}p_{\theta}(\bm{x}_{t-1}|\bm{x}_t),\\\text{with\,\,} \,  
p_{\theta}(\bm{x}_{t-1}|\bm{x}_t)=\mathcal{N}(\bm{\mu}_{\theta}(\bm{x}_t,t),\bm{\Sigma}_{\theta}(\bm{x}_t,t)).
\end{align*}
Here, following \cite{ho2020denoising}, $\bm{\mu}_{\theta}$ is estimated by a Unet \cite{ronneberger2015u} with trainable parameters $\theta$, and $\bm{\Sigma}_{\theta}(t) =\bm{\Sigma}(t) = \frac{1-\alpha_{t-1}}{1-\alpha_t} \beta_t \textbf{I}$ is fixed. Variational inference is used to achieve a tractable loss function and the variational lower bound (VLB) is derived as 
\begin{align*}\mathcal{L}_{VLB}=-log(p_{\theta}(\bm{x}_0)) \\ +D_{KL}(q(\bm{x}_{1:T}|\bm{x}_0)||p_{\theta}(\bm{x}_{1:T}|\bm{x}_0)).
\end{align*}
which can be reformulated to 
\begin{align*} 
\mathcal{L}_{simple} = ||\bm{\epsilon} - \bm{\epsilon}_{\theta}(\bm{x}_t,t)||^2
\end{align*}
by applying simplifications and by conditioning the denoising step on $\bm{x}_0$, as shown in \cite{ho2020denoising}.
In our work, instead of predicting the noise $\bm{\epsilon}$ we perform the equivalent task of directly estimating $\bm{x}^{rec}_{0} = \bm{x}_t - \bm{\epsilon}$. Hence, we derive our loss function as 
\begin{equation*}
\mathcal{L}_{rec} = |\bm{x}_0 - \bm{x}^{rec}_{0}|. 
\end{equation*}
Typically, to generate new images with DDPMs, the backward step is applied step-wise to gradually denoise a random noise vector. For the given UAD task, we do not aim to generate new images but to estimate healthy brain anatomy given an input image. Therefore, we directly estimate $\bm{x}^{rec}_{0}$ given $\bm{x}_t$ at test time as it is done in \cite{BehrendtpDDPM}. The time step $t_{test}<T$ controls the level of noise to remove from $\bm{x}_t$ at test time. Optionally, to become agnostic to the noise magnitude, we use an ensemble of different values $t_{test}=[250,500,750]$ and average the reconstructions of each noise level, similar to \cite{graham2022denoising}. 
\subsection{Conditioned DDPMs (cDDPMs)}
A general depiction of our conditioning approach is provided in Figure \ref{fig:graphical abstract}.
Formally, we condition the backward process of DDPMs on a context vector $\bm{c}$ as follows
\begin{align*}
    \bm{x}_{0}^{rec} \sim p(\bm{x}_T)\prod\nolimits^T_{t=1}p_{\theta}(\bm{x}_{t-1}|\bm{x}_t,\bm{c}),\\ \text{with\,\,} \,  
p_{\theta}(\bm{x}_{t-1}|\bm{x}_t,\bm{c})=\mathcal{N}(\bm{\mu}_{\theta}(\bm{x}_t,t,\bm{c}),\bm{\Sigma}(t)).
\end{align*}
We use an image encoder $F_{enc}$ to achieve a latent representation $\bm{c} = F_{enc}(\bm{x}_0)$ of the input image $\bm{x}_0$ where $\bm{c} \in \mathbb{R}^d$ with $d$ as conditioning dimension. \\
To integrate the context vector $\bm{c}$, we manipulate the denoising Unet of the DDPM. Therefore, we individually adapt the features $\bm{f}_i \in \mathbb{R}^{H_i,W_i,C_i}$ at each level of the denoising Unet based on $\bm{c}$ where $H_i$, $W_i$ and $C_i$ are the respective feature map dimensions. 
To achieve this, we adapt the time step conditioning of DDPMs as follows. First, the time step is encoded using a sinusoidal position embedding. Next, $t$ is projected to a vector $\bm{c}_t \in \mathbb{R}^{d}$ by a multi-layer perceptron (MLP I). Subsequently, we concatenate the context vector $\bm{c}$ and the time step vector $\bm{c}_t$, resulting in a conditioning vector $\bm{c}' \in \mathbb{R}^{2 \cdot d}$.
Finally, $\bm{c}'$ is projected to $\bm{c}_{proj}' \in \mathbb{R}^{2\cdot C_i}$ by another multi-layer perceptron (MLP II) at each feature level $i$. The vector $\bm{c}_{proj}'$ is then split into half, where the first and last $C_i$ elements resemble the scaling factor $\gamma$ and the shift value $\beta$.  Inspired by \cite{perez2018film}, the variables $\gamma$ and $\beta$ are used to transform the individual feature maps as $\bm{f}_i' = \bm{f}_i \ast (\gamma + 1) + \beta$ in each residual block. This transformation adaptively scales and shifts the feature maps at each level of the denoising UNet, based on the context vector $\bm{c} = F_{enc}(\bm{x}_0)$, which encodes relevant information from the input image. The purpose of this transformation is to allow the model to dynamically adjust feature representations in response to both the image features and the conditioning information. By modulating the feature maps through context-based scaling and shifting, the model can more effectively preserve critical details of the clean target image while denoising the noisy input.\\
Optionally, to achieve a meaningful starting point for the calculation of the context vector $\bm{c} = F_{enc}(\bm{x}_0)$, we pre-train the feature extraction of the image encoder $F_{enc}$ which is described in the next section.   
\subsection{Pre-Training}
\begin{figure}[h!]
    \centering
    \includegraphics[width=.5\linewidth]{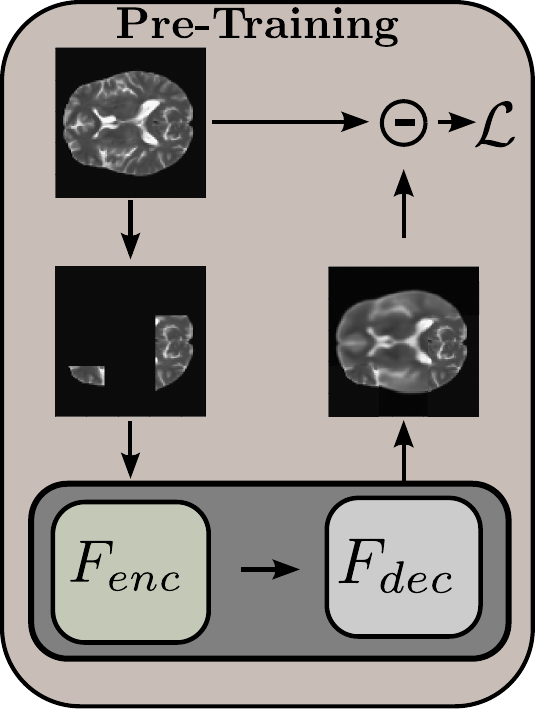}
    \caption{Overview of the pre-training strategy. Random patches are erased from the input image. $F_{enc}$ is used to derive a latent representation and $F_{dec}$ is used to reconstruct the removed patches. After pre-training, the encoder $F_{enc}$ is then fine-tuned along with the \textit{DDPM} in the main training stage to condition the denoising process.}
    \label{fig:graphical abstract pre}
\end{figure}

We utilize a generative pre-training strategy for $F_{enc}$. More precisely, we utilize masked pre-training where typically transformer-based AEs are trained to reconstruct an image where a significant fraction of patches are masked out \cite{he2022masked}. We adopt the SparK framework \cite{tian2023designing}, where sparse convolutions and hierarchical features are used to enable the masked pre-training for CNNs. We pre-train the encoder with the same healthy training set as the main training task to learn the general feature representations required to capture important information from the MRI scans. After the pre-training stage, we discard the decoder and only use $F_{enc}$ and fine-tune it along with the denoising Unet during the main training stage of the cDDPM.  A schematic description of the pre-training stage is provided in Figure \ref{fig:graphical abstract pre}.
\section{Experimental Setup} \label{sec:expsetup}
\subsection{Data Sets}
Following the principle of UAD, we train our models for the reconstruction task on healthy data only (IXI). At test time, we evaluate the models' anomaly detection ability on unhealthy test sets of various pathologies (BraTS21, ATLAS, MSLUB, WMH). We provide an overview of all available MRI scanner details for the different data sets in Table \ref{tab:app:scanner}.
\subsubsection{Training Data}
We use the publicly available IXI data set\footnote{https://brain-development.org/ixi-data set/} as our healthy reference data set for training. This data set includes 560 3D brain MRI scans collected from three medical facilities. Of the training data, 158 samples are set aside for testing, while the remaining data is divided into 5 folds, each containing 358 training samples and 44 validation samples for cross-validation. \\ Note that only the healthy training and validation data is used during pre-training and training stage.
\subsubsection{Evaluation Data}
For evaluation, we utilize four different publicly available data sets that contain different types of pathologies and the corresponding manual expert annotations: 
\begin{enumerate}
    \item Multimodal Brain Tumor Segmentation Challenge 2021 (BraTS21) \cite{baid2021rsna,menze2014multimodal,bakas2017advancing}
    \item Multiple Sclerosis data set from the University Hospital of Ljubljana (MSLUB) \cite{lesjak2018novel}
    \item Anatomical Tracings of Lesions After Stroke v2.0 (ATLAS) \cite{liew2022large}
    \item White Matter Hyperintensity (WMH) \cite{kuijf2019standardized}
\end{enumerate}

The BraTS21 data set includes 2040 3D brain routine MRI scans of patients with glioma with a pathologically confirmed diagnosis. Accompanying the MRI scans, annotations from expert neuroradiologists are provided for 1251 scans that delineate tumor sub-regions as categorical masks. We fuse all sub-regions to obtain a binary segmentation mask to evaluate the anomaly detection task. All scans are available as T1-weighted volumes with and without contrast enhancement (T1-CE, T1) and T2-weighted or T2 fluid-attenuated inversion recovery (T2, FLAIR) volumes.   
The MSLUB data set includes 3D brain MRI scans of 30 patients with multiple sclerosis (MS) lesions. For each patient, along with the T1, T2 and FLAIR MRI scans, ground truth annotations are available derived based on multi-rater consensus.
The ATLAS data set consists of 655 T1-weighted MRI scans of stroke patients collected from 44 research cohorts. The stroke lesions are annotated by domain experts and binary segmentation masks are provided.
The WMH data set consists of 60 MRI scans of patients with white matter hyperintensities from three different institutions and scanner types. WMH segmentation masks are derived from the consensus of two expert radiologists. 
The datasets contain images acquired with different MRI parameters. While BraTS21 and MSLUB include T1,T2, and FLAIR data, WMH contains only T1 and FLAIR data, and ATLAS only T1 data. The IXI data set used for training contains T1 and T2 data. Hence, we train the models separately on T1 and T2 data, and evaluate on the respective data sets.  \\
\subsection{Pre-Processing}
We pre-process the images according to established pre-processing strategies for UAD in brain MRI \cite{baur2021autoencoders}. First, we resample all MRI scans to the isotropic resolution of $1\mathrm{\,mm}\times 1\mathrm{\,mm} \times 1\mathrm{\,mm}$ using cubic spline interpolation. Second, we register all MRI scans to the SRI24-Atlas. Third, we remove the skull from the MRI scans by skull stripping with HD-BET \cite{isensee2019automated}. Subsequently, we crop each brain scan using its corresponding brain mask, removing unnecessary background while preserving relevant brain tissue. We then apply N4 bias field correction \cite{tustison2010n4itk}. To standardize image sizes, we identify the largest brain volume in our dataset, add a safety margin, and pad all images to a unified size of $192 \times 192 \times 160$. Finally, to save computational resources, we reduce volume resolution by a factor of two and remove the 15 top and bottom slices parallel to the transverse plane, leading to a final resolution of $96 \times 96 \times 50$\, voxels.
\subsection{Post-Processing}
At test time, we derive a binary segmentation map from the residual map $\bm{R}=|\bm{x}_0-\bm{x}_0^{rec}|$ where regions of high residuals indicate anomalies. To binarize $\bm{R}$, we first apply the following post-processing steps that are commonly used in the field of UAD in brain MRI \cite{baur2021autoencoders,kascenas2021denoising,BehrendtpDDPM,zimmerer2018context}. First, a median filter with a kernel size of $K=5\times5\times5$ is applied to smooth the residual map and to remove false positives. Subsequently, we perform brain mask eroding for 3 iterations. This step is mainly applied to filter out residuals due to poor reconstructions at sharp edges near the brain mask \cite{baur2021autoencoders}. 
After binarization, we use connected component filtering to remove areas that include less than 7 voxels. This post-processing step removes false-positive predictions smaller than the anomalies considered in this study. In the supplemental material, we systematically compare the post-processing strategies for each data set.

\subsection{Implementation Details}
This study compares our proposed cDDPM method with multiple established baselines for UAD in brain MRI. The baselines include \textit{AE}, \textit{VAE}  \cite{baur2021autoencoders}, \textit{SVAE} \cite{behrendt2022capturing}, Reverse Anoamly \textit{RA} \cite{bercea2023generalizing}, \textit{PHANES} \cite{Bercea.2023MICa} and denoising AEs \textit{DAE} \cite{kascenas2021denoising}. We also compare with simple thresholding \textit{Thresh} \cite{meissen2022challenging}. Furthermore, we compare the feature-based reverse distillation method \textit{RD} \cite{Deng.2022}, Feature-Autoencoders \textit{FAE} \cite{Meissen.2022FAE}, the Encoder-Decoder Contrast method \textit{EDC} \cite{Guo.2023} and the self-supervised Poisson Image Interpolation \textit{PII} \cite{Tan.2021}. For a direct comparison, we also include \textit{DDPM} \cite{wyatt2022anoddpm} and patched DDPMs \textit{pDDPM} \cite{BehrendtpDDPM} as a counterpart to our proposed method.\\
We adapt the baseline implementations by tuning hyper-parameters based on the validation set to improve training stability and performance. We set $\beta_{VAE}$ to 0.001 for \textit{VAE} and \textit{SVAE}. 
For \textit{EDC}, we use a lr of 1e-5 and 5e-5 for the encoder and decoder, respectively. \textit{RA} and \textit{PHANES} are trained with $\beta_{rec}=16$ for the Soft-Intro VAE. For \textit{PHANES}, we use a masking threshold of $0.15$ and $0.10$ for training and testing, respectively.\\ 
For \textit{DDPM}, \textit{pDDPM} and cDDPM, we use structured simplex noise, which has been shown to improve the UAD performance on MRI images \cite{wyatt2022anoddpm}.
Furthermore, we uniformly sample $t \in [1,T]$ with $T=1000$ during training. At test time, we either use a fixed value of $t_{test}=\frac{T}{2}=500$ or an ensemble of different values $t_{test}=[250,500,750]$ and average the individual reconstructions of each noise level. The denoising network for all DDPM-based approaches is an Unet similar to \cite{dhariwal2021diffusion}, with channel dimensions of [128, 256, 256].
As encoder network $F_{enc}$, we utilize a ResNet-backbone with a fully connected layer to match the target dimension of $\bm{c}\in \mathbb{R}^{d}$ with $d=128$ as conditioning dimension. We evaluated several conditioning dimensions, $d \in \{32, 64, 128, 256, 512\}$, and observed that the model achieved optimal performance when $d=128$. A mask-out ratio of 65 $\%$ is used during pre-training.
For data augmentation, we utilize random -blur (p=0.25), -bias (p=0.25), -gamma (p=0.5) and -ghosting (p=0.5) from the torchio library \cite{PEREZGARCIA2021106236}.
We train for a maximum of 1600 epochs on NVIDIA V100 (32GB) GPUs, using Adam as an optimizer, a learning rate of $1e-4$, and a batch size of 32. The best model checkpoint, as determined by performance on the validation set, is used for testing. The volumes are processed slice-wise, uniformly sampling slices with replacement during training and iterating over all slices to reconstruct the full volume at test time. We implement all models in Pytorch (v0.10).

\subsection{Experiments and Evaluation}
We conduct various experiments to assess the individual features of our proposed cDDPMs. First, we evaluate the reconstruction quality for both healthy and unhealthy structures. Second, we investigate the generalization to real and simulated domain shifts between the training and test data. Lastly, we assess the final segmentation performance. For all experiments, we compare our cDDPMs to established baseline models on different data sets. The following subsections provide the detailed experimental settings for each individual evaluation.
\subsubsection{Reconstruction Quality} To evaluate the overall reconstruction quality, we utilize the held-out test set of the healthy IXI data set and calculate similarity metrics between input and reconstruction. We consider the Structural Similarity Index Measure (SSIM) \cite{SSIM}, the Peak Signal To Noise Ratio (PSNR) and the Learned Perceptual Image Patch Similarity (LPIPS) as metrics to asses the reconstruction quality. For the feature-based LPIPs metric \cite{zhang2018unreasonable}, features are extracted by a ResNet-based network, pre-trained on 3D medical data \cite{chen2019med3d}. Furthermore, we report the reconstruction error ($l1$-error) for the healthy data set. For UAD, only healthy anatomy should be reconstructed. Hence, it is interesting to consider the $l1$-error of healthy and unhealthy anatomy separately, given the unhealthy evaluation data sets. Therefore, we calculate the $l1$-error for both healthy and unhealthy anatomy, as indicated by the annotation masks and calculate an $l1$-ratio as follows:
\[
l1\text{-ratio} = \frac{l1_{unhealthy}}{l1_{healthy}}.
\]
A higher value for the $l1$-ratio indicates that the model successfully reconstructs the healthy anatomy while struggling to reconstruct the unhealthy parts of the input, and vice versa.

\subsubsection{Domain Adaptation and Intensity Alignment} We investigate our proposed approach's generalization to domain shifts and the capability of adequately reconstructing the local intensity patterns of a given input image. We utilize data sets from different domains. For training, we utilize in-domain data from the IXI data set. As out-of-domain data, we utilize the BraTS21 data set. Notably, we only consider the content of the BraTS21 data set that is marked as healthy by the provided annotation masks, thereby ensuring the evaluation of domain shifts regarding scanner and brain diversity, not domain shifts arising from unhealthy brain MRI structures. Furthermore, we simulate different contrast levels ranging from $cl \in [0.3, 0.7, 1.0, 1.5, 2.0]$. The images of different contrast levels are derived by potentiating the gray values by the respective contrast level, i.e., $\bm{x}_0^{cl=2} = \bm{x}_0^2$. 
To evaluate the effect of the conditioning mechanism, we utilize the IXI data set and simulate different levels of conditioning information to investigate the reconstructions qualitatively. Therefore, we alter the available information in the image fed to the image encoder to condition the cDDPM. To achieve this, we crop the conditioning image at a given width of 50$\%$  and 100$\%$ where 100$\%$ indicates that the entire input image is used as the conditioning image.  In addition to the qualitative evaluation of reconstructed, simulated data, we quantitatively assess the domain shift across input and reconstruction by comparing their intensity histograms. Therefore, we first calculate and plot the histograms. We partition the intensity values into 500 bins and divide the raw count by the total number of counts and the bin width. For a quantitative analysis of the distance between the intensity distributions, we calculate the Kullback-Leibler Divergence (KLD) as follows:
\begin{align*}
\text{{KLD}} = \left[-\sum_{i} p_{input} \log(p_{input}) \right] - \\ \left[ -\sum_{i} p_{reconstruction} \log(p_{reconstruction}) \right]
\end{align*}
where $p = [p_1, p_2, \ldots, p_n]$ represents each intensity distribution. 

\subsubsection{Segmentation Performance} To assess the segmentation performance for the UAD task, we utilize all unhealthy test sets and consider two different segmentation metrics. First, we report the best possible Dice score ($\lceil$DICE$\rceil$).
The formula for the Dice score is given by:
\[
\text{{DICE}} = \frac{{2 \cdot |A \cap B|}}{{|A| + |B|}}
\]
where \(A\) and \(B\) are the predicted anomaly map and the ground truth annotation, respectively.
The best possible $\lceil$DICE$\rceil$ is estimated by a greedy search of the threshold that optimizes the Dice score, similar to \cite{Lagogiannis.2023}.
\\
Second, we calculate the Area Under Precision-Recall Curve (AUPRC) as follows:
\[
\text{{AUPRC}} = \sum_{r} (R(r) - R(r-1)) \cdot P(r).
\]
Here, \(R(r)\) represents the recall at a given threshold or rank \(r\), and \(P(r)\) represents the precision at the corresponding recall \(R(r)\). The sum is taken over all thresholds or ranks \(r\) at which the precision and recall are computed.
\\
\subsubsection{Statistical Testing} To conduct significance tests, we utilize the MLXtend library's permutation test \cite{raschkas_2018_mlxtend} with 10,000 rounds of permutations and a significance level of $\alpha=5\%$. This test computes the mean difference of the considered scores of two models for each permutation, and the resulting p-value is computed by counting the number of times the mean differences were equal to or greater than the sample differences, divided by the total number of permutations.

\section{Results} \label{sec:results}
\begin{table*}[!t]
\centering
\caption{Comparison of the reconstruction quality of the different models with the best results highlighted in bold. The asterisk * denotes superior performance with statistical significance compared to all other models ($p<0.05$). For all metrics, the mean $\pm$ standard deviation across the different folds are reported. The arrows $\uparrow$ and $\downarrow$ indicate that higher and lower values are favorable, respectively. The $l1$-ratio is derived by dividing the $l1$-error of unhealthy anatomy by the $l1$-error of healthy anatomy. DDPM-based models are evaluated by ensembling different values for $t_{test}=[250,500,750]$}
\resizebox{\linewidth}{!}{
\begin{tabular}{lcccccccc}
\toprule
 & \multicolumn{4}{c}{IXI (T2)} & BraTS21 (T2) & MSLUB (T2) & ATLAS (T1) & WMH (T1)  \\
 \cmidrule(l){2-5} \cmidrule(l){6-6} \cmidrule(l){7-7}  \cmidrule(l){8-8} \cmidrule(l){9-9}
 \textbf{Model} &                \textbf{SSIM} $\uparrow$  &             \textbf{PSNR} $\uparrow$ &          \textbf{LPIPS} (e-3) $\downarrow$ &         $\bm{l1\textbf{-error}}$ (e-3) $\downarrow$ &  $\bm{l1\textbf{-ratio}}$ $\uparrow$ &  $\bm{l1\textbf{-ratio}}$ $\uparrow$&  $\bm{l\textbf{-ratio}}$ $\uparrow$&  $\bm{l1\textbf{-ratio}}$ $\uparrow$\\
\midrule
   \textit{VAE} \cite{baur2021autoencoders} & 74.98 $\pm$ 0.54 & 23.38 $\pm$ 0.14 &  4.03 $\pm$ 0.50 & 32.32 $\pm$ 0.64 & 3.52 $\pm$ 0.08 & 2.92 $\pm$ 0.06 &   4.43 $\pm$ 0.03 & 2.36 $\pm$ 0.04 \\
 \textit{SVAE} \cite{behrendt2022capturing}& 77.87 $\pm$ 0.15  & 23.94 $\pm$ 0.06 & 3.31 $\pm$ 0.24 & 29.08 $\pm$ 0.16 &  3.90 $\pm$ 0.05 & 3.13 $\pm$ 0.05 &  3.38 $\pm$ 0.11 & 2.07 $\pm$ 0.01 \\
   \textit{AE} \cite{baur2021autoencoders} & 76.11 $\pm$ 0.27 & 23.41 $\pm$ 0.14 & 3.19 $\pm$ 0.54 & 31.67 $\pm$ 0.41 &  3.84 $\pm$ 0.17  & 3.26 $\pm$ 0.18 &    4.40 $\pm$ 0.07 & 2.36 $\pm$ 0.04 \\
\textit{RA} \cite{bercea2023generalizing}&75.46 $\pm$ 0.35 & 23.92 $\pm$ 0.23 & 2.18 $\pm$ 0.41 &         34.36 $\pm$ 1.43 &                 3.10 $\pm$ 0.16 &              2.56 $\pm$ 0.11 & 3.93 $\pm$ 0.25 &            2.42 $\pm$ 0.19\\
   \textit{PHANES} \cite{Bercea.2023MICa}& 69.04 $\pm$ 1.23 & 21.39 $\pm$ 0.32 & 1.08 $\pm$ 0.09 &          38.70 $\pm$ 1.74 &                3.54 $\pm$ 0.13 &              2.73 $\pm$ 0.07 & 4.01 $\pm$ 0.07 &             2.20 $\pm$ 0.05 \\
  \textit{DAE} \cite{kascenas2021denoising}   & \textbf{98.69 $\pm$ 0.15}* & \textbf{36.69 $\pm$ 0.38}* & 0.14 $\pm$ 0.01 &  \textbf{8.14 $\pm$ 0.17}* & 7.17 $\pm$ 0.63 & 2.69 $\pm$ 0.15 &   4.51 $\pm$ 0.15 & 2.99 $\pm$ 0.14 \\

  \midrule
   \textit{DDPM} \cite{wyatt2022anoddpm} & 93.96 $\pm$ 0.37 & 31.79 $\pm$ 0.26 & 0.49 $\pm$ 0.14 & 14.29 $\pm$ 0.32 & 6.16 $\pm$ 0.53 & 3.37 $\pm$ 0.24 &    5.00 $\pm$ 0.23 & \textbf{3.16 $\pm$ 0.15} \\
\textit{pDDPM} \cite{BehrendtpDDPM} & 96.62 $\pm$ 0.25 & 34.58 $\pm$ 0.39 & \textbf{0.09 $\pm$ 0.04}* &   9.70 $\pm$ 0.43 & 7.16 $\pm$ 0.15 & 4.34 $\pm$ 0.13 &   5.58 $\pm$ 0.28 &  3.00 $\pm$ 0.16 \\
cDDPM (Ours) &  96.80 $\pm$ 0.19  & 34.87 $\pm$ 0.23 & 0.11 $\pm$ 0.05 &  9.68 $\pm$ 0.16 & \textbf{7.43 $\pm$ 0.17} & \textbf{4.49 $\pm$ 0.18} &\textbf{5.69 $\pm$ 0.27} & 3.12 $\pm$ 0.08 \\
\bottomrule
\end{tabular}}
\label{tab:recoresults}
\end{table*}
\begin{figure*}[!h]
    \centering
    \includegraphics[width=\linewidth]{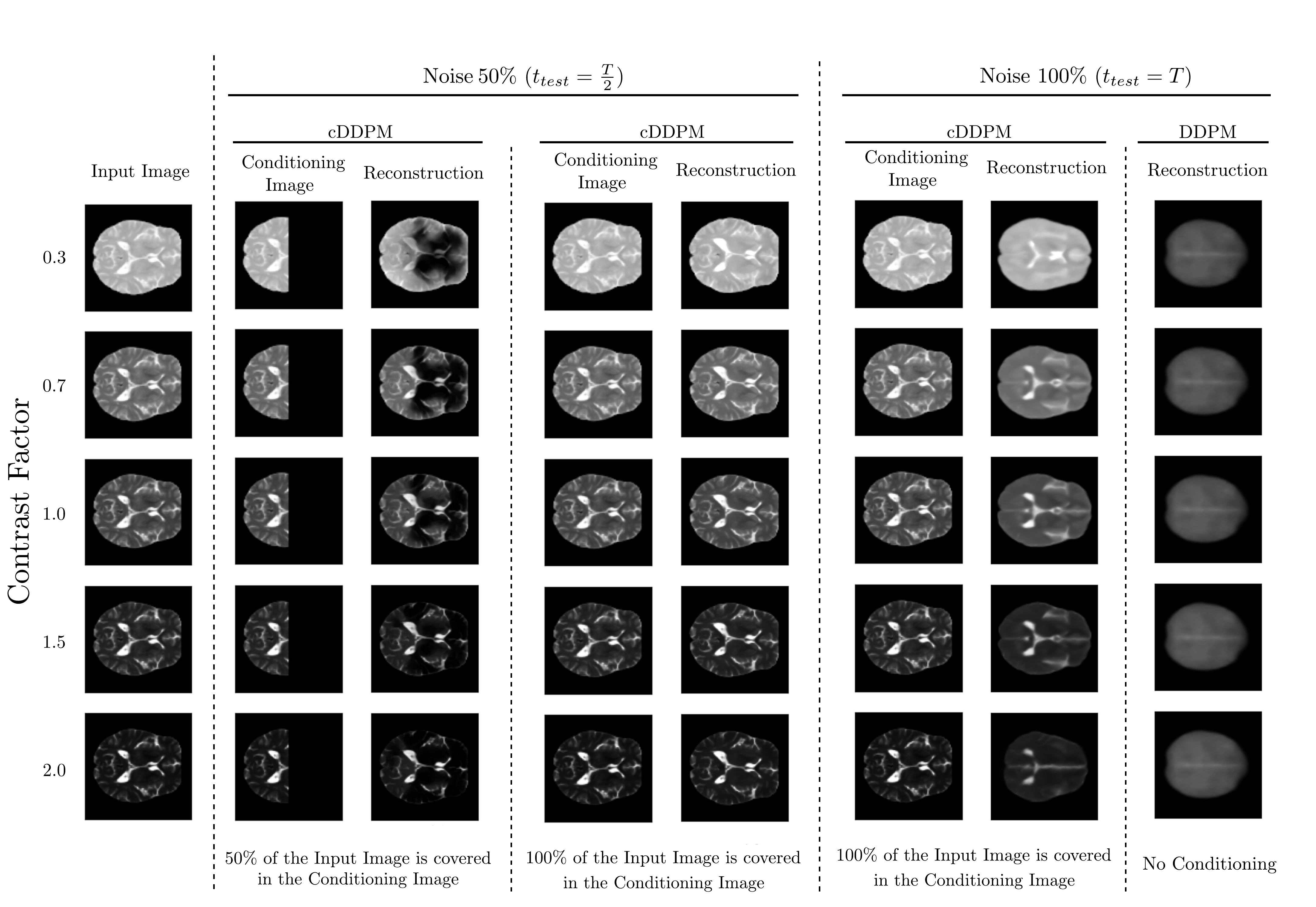}

    \caption{Simulating the conditioning effect of the cDDPM for 50 $\%$ of noise and 100$\%$ noise in the input image. In the first block, the input image that is fed to the \textit{DDPM} or cDDPM is shown. In the second block, the reconstructions of cDDPM for different conditioning inputs are shown when a noise level of 50$\%$ is applied. In the third block, the reconstructions of cDDPMs and DDPMs are compared at a noise level of 100$\%$. From top to bottom, the contrast level of the conditioning and input image is increased, respectively, for all columns.}
    \label{fig:cond_eval}
\end{figure*}

\begin{figure*}[!h]
    \centering
    \includegraphics[width=.9\linewidth]{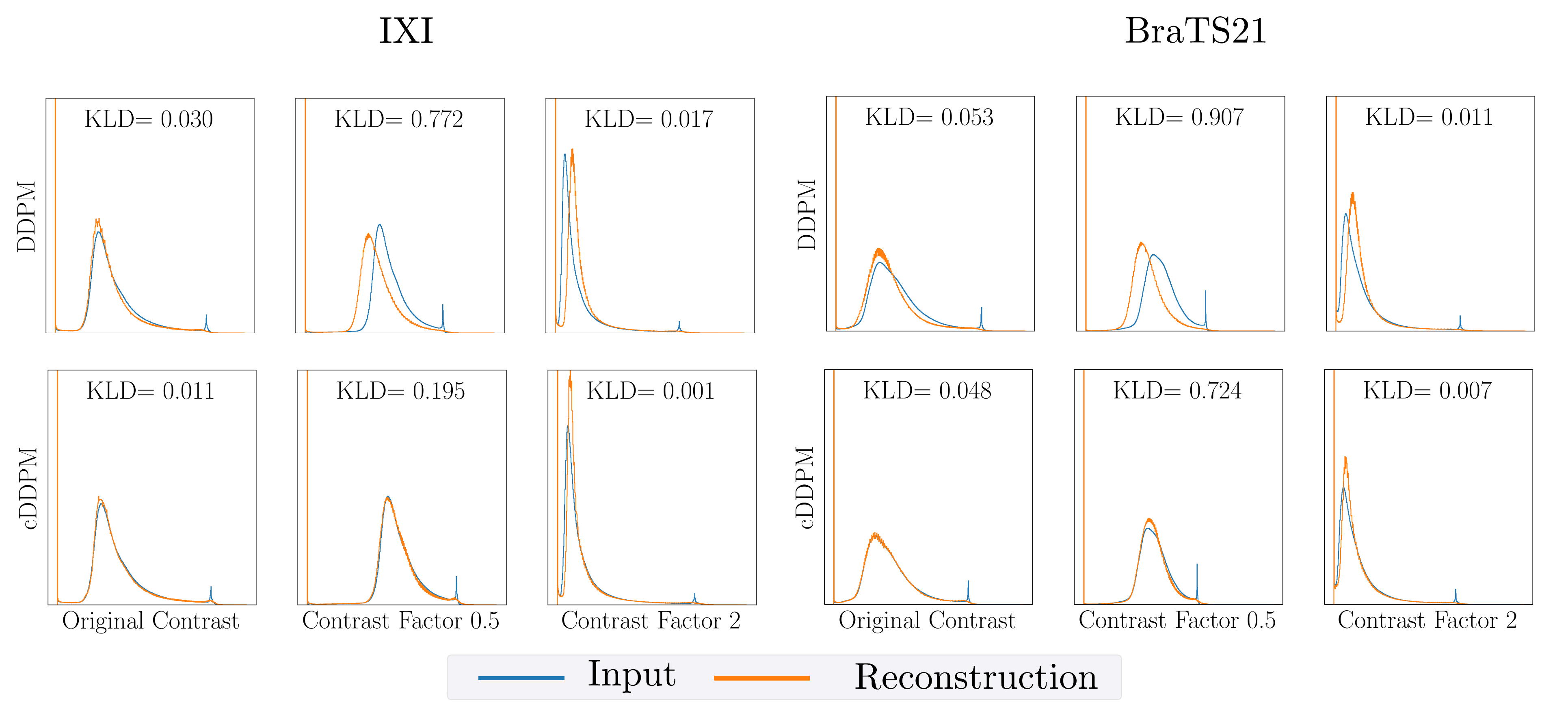}
    
    \caption{Comparison of the histograms for input-reconstruction pairs of the healthy IXI (left) and the unhealthy BraTS21 (right) data set with original and augmented contrast. The top row shows the baseline \textit{DDPM} without conditioning and the bottom row our proposed cDDPM with conditioning. The Kullback-Leibler divergence (KLD) for both histograms is indicated within each plot (lower values are preferable). Both models are evaluated by ensembling different values for $t_{test}=[250,500,750]$. }
    \label{fig:histos_ixi_brats}
\end{figure*}

\begin{table*}[h!]
\centering
\caption{Comparison of the evaluated models with the best results highlighted in bold. The asterisk * denotes superior performance with statistical significance compared to all other models ($p<0.05$). For all metrics, the mean $\pm$ standard deviation across the different folds are reported. A checkmark at SSL denotes that a pre-trained encoder is used. A checkmark at ENS denotes the ensembling of different values for $t_{test}=[250,500,750]$. Otherwise, a fixed value of $t_{test}=500$ is used for DDPM-based models. }
\resizebox{\linewidth}{!}{

\begin{tabular}{lcccccccccc}
\toprule
 & \multicolumn{2}{c}{Modification} & \multicolumn{2}{c}{BraTS21 (T2)}  & \multicolumn{2}{c}{MSLUB (T2)} & \multicolumn{2}{c}{ATLAS (T1)} & \multicolumn{2}{c}{WMH (T1)} \\
\cmidrule(l){2-3} \cmidrule(l){4-5} \cmidrule(l){6-7} \cmidrule(l){8-9} \cmidrule(l){10-11} 
\textbf{Model} & \textbf{ENS} & \textbf{SSL} & \textbf{$\lceil$DICE$\rceil$ [$\%$]} & \textbf{AUPRC [$\%$]} & \textbf{$\lceil$DICE$\rceil$ [$\%$]} & \textbf{AUPRC [$\%$]} & \textbf{$\lceil$DICE$\rceil$ [$\%$]} & \textbf{AUPRC [$\%$]}& \textbf{$\lceil$DICE$\rceil$ [$\%$]} & \textbf{AUPRC [$\%$]} \\
\midrule
            \textit{Thresh} \cite{meissen2022challenging} &&&      30.26  & 20.27  &    7.65  & 4.23 &       4.66 &   1.71  & 10.32 & 4.72\\
              \textit{VAE} \cite{baur2021autoencoders} &  &  &  33.12 $\pm$ 1.12 & 25.74 $\pm$ 1.37 &   8.10 $\pm$ 0.18 &  4.48 $\pm$ 0.18 &   15.63 $\pm$ 0.73 &  11.44 $\pm$ 0.5 &   7.60 $\pm$ 0.28 &  3.86 $\pm$ 0.40 \\
              \textit{SVAE} \cite{behrendt2022capturing} &  &  & 36.43 $\pm$ 0.36 & 30.3 $\pm$ 0.45 & 8.55 $\pm$ 0.11 & 4.8 $\pm$ 0.09 &  10.32 $\pm$ 0.53 & 6.84 $\pm$ 0.44 & 7.18 $\pm$ 0.07 & 2.97 $\pm$ 0.06  \\
               \textit{AE} \cite{baur2021autoencoders} &  &  &  36.04 $\pm$ 1.73 &  28.8 $\pm$ 1.72 &  9.65 $\pm$ 0.97 &   5.71 $\pm$ 0.80 &    14.04 $\pm$ 0.6 & 10.16 $\pm$ 0.53 &  7.34 $\pm$ 0.08 & 3.43 $\pm$ 0.14 \\
                 \textit{DAE} \cite{kascenas2021denoising} &  &  & 48.82 $\pm$ 3.68 & 49.38 $\pm$ 4.18 &  7.57 $\pm$ 0.61 &  4.47 $\pm$ 0.69 &   15.95 $\pm$ 0.69 & 13.37 $\pm$ 0.62 & 12.02 $\pm$ 1.01 & 8.54 $\pm$ 1.02 \\
                 \textit{RA} \cite{bercea2023generalizing} &&& 16.75 $\pm$ 0.51 & 9.98 $\pm$ 0.43 & 3.96 $\pm$ 0.03 & 1.92 $\pm$ 0.04 & 12.21 $\pm$ 0.98 & 8.75 $\pm$ 0.93 & 6.04 $\pm$ 0.45 & 3.15 $\pm$ 0.31 \\
                \textit{PHANES} \cite{Bercea.2023MICa} &&&  28.42 $\pm$ 0.91 & 21.29 $\pm$ 1.06 & 6.11 $\pm$ 0.27 & 2.98 $\pm$ 0.07 & 17.62 $\pm$ 0.41 & 13.81 $\pm$ 0.48 & 7.55 $\pm$ 0.17 & 3.87 $\pm$ 0.13 \\

              \textit{RD} \cite{Deng.2022} &&&  32.57 $\pm$ 0.15 & 27.11 $\pm$ 0.15 &  6.48 $\pm$ 0.20 & 3.32 $\pm$ 0.06 &   19.69 $\pm$ 0.26 & 15.51 $\pm$ 0.20 &  7.48 $\pm$ 0.10 & 3.89 $\pm$ 0.07 \\
               \textit{FAE} \cite{Meissen.2022FAE} &&& 44.59 $\pm$ 2.19 & 43.63 $\pm$ 0.47 & 6.85 $\pm$ 0.65 & 3.85 $\pm$ 0.08 &   17.76 $\pm$ 0.16 & 13.91 $\pm$ 0.10 & 8.81 $\pm$ 0.38 & 4.77 $\pm$ 0.26   \\
                \textit{EDC} \cite{Guo.2023} &&&36.66 $\pm$ 3.03 & 30.47 $\pm$ 4.25 & 7.23 $\pm$ 0.29 & 3.88 $\pm$ 0.4 &   18.67 $\pm$ 1.02 & 14.34 $\pm$ 0.86 & 8.62 $\pm$ 0.47 & 4.67 $\pm$ 0.37 \\

               \textit{PII} \cite{Tan.2021} &&&  40.83 $\pm$ 2.18 & 36.52 $\pm$ 2.66 & 9.46 $\pm$ 0.43 & 5.22 $\pm$ 0.37& 9.73 $\pm$ 1.89& 7.31 $\pm$ 1.64 & 6.59 $\pm$ 1.87 & 3.36 $\pm$ 1.03 \\

            \midrule
             \textit{DDPM} \cite{wyatt2022anoddpm}  &  &  & 49.43 $\pm$ 1.94 &  50.00 $\pm$ 2.13 &  9.63 $\pm$ 1.33 &  6.81 $\pm$ 1.54 &   17.57 $\pm$ 1.05 &  15.64 $\pm$ 0.90 & 11.56 $\pm$ 0.93 & 8.65 $\pm$ 0.87 \\
         \textit{DDPM} \cite{wyatt2022anoddpm} & \cmark &  & 50.27 $\pm$ 2.67 & 50.61 $\pm$ 2.92 &  9.71 $\pm$ 1.29 &  6.27 $\pm$ 1.58 &   20.18 $\pm$ 0.58 & 17.77 $\pm$ 0.47 & \textbf{12.06 $\pm$ 0.97} & 8.89 $\pm$ 0.89 \\
        \textit{pDDPM} \cite{BehrendtpDDPM}&  &  &53.25 $\pm$ 0.50 & 54.73 $\pm$ 0.52 &  12.40 $\pm$ 0.36 & 10.14 $\pm$ 0.44 &    19.20 $\pm$ 0.45 & 17.31 $\pm$ 0.34 &  10.14 $\pm$ 0.50 & 7.78 $\pm$ 0.56 \\
    \textit{pDDPM} \cite{BehrendtpDDPM}& \cmark &  & 53.61 $\pm$ 0.51 & 55.08 $\pm$ 0.54 &  12.83 $\pm$ 0.40 & 10.02 $\pm$ 0.36 &   19.92 $\pm$ 0.24 &  17.84 $\pm$ 0.10 & 10.13 $\pm$ 0.53 & 7.52 $\pm$ 0.56 \\
\midrule
cDDPM (Ours) &  &  &  54.49 $\pm$ 1.63 & 56.81 $\pm$ 1.96 & 12.79 $\pm$ 1.08 & 10.07 $\pm$ 1.07 &    22.6 $\pm$ 1.67 & 20.65 $\pm$ 1.52 & 11.21 $\pm$ 0.54 & 9.05 $\pm$ 0.56 \\
      cDDPM (Ours) &  & \cmark & 55.67 $\pm$ 1.05 & 58.14 $\pm$ 1.28 & 13.52 $\pm$ 0.91 & 10.89 $\pm$ 1.08 &    22.66 $\pm$ 1.20 & 20.85 $\pm$ 1.28 &  11.15 $\pm$ 0.8 &  9.03 $\pm$ 0.90 \\
  cDDPM (Ours) & \cmark & \cmark & \textbf{56.30 $\pm$ 1.25*} & \textbf{58.82 $\pm$ 1.56*}& \textbf{14.04 $\pm$ 1.16} & \textbf{10.97 $\pm$ 1.17} &    \textbf{24.22 $\pm$ 1.10*} & \textbf{22.22 $\pm$ 1.15*} & 11.59 $\pm$ 0.93 & \textbf{9.26 $\pm$ 1.07} \\
\bottomrule
\end{tabular}
}
\label{tab:mainresults}
\end{table*}

\begin{figure*}[!h]
\centering
\includegraphics[width=\linewidth]{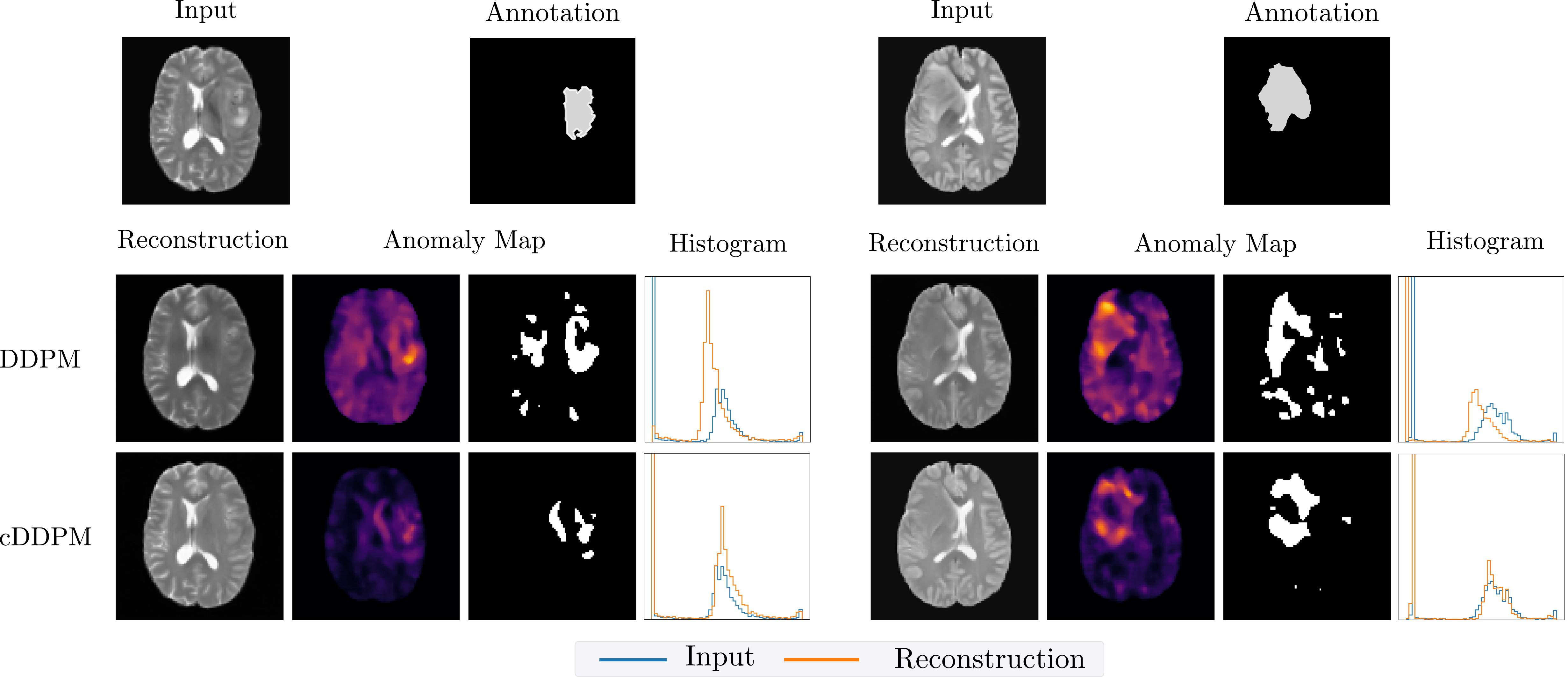} 
\caption{Exemplary reconstructions and anomaly maps for DDPMs (second row) and cDDPMs (third row). The input and the corresponding ground truth annotation are provided in the first row. For each case, the reconstruction, the anomaly map and the histograms of intensity values in input and reconstruction are shown.  Note that for histogram calculation, only healthy areas are considered. For visualization purposes, we provide segmentation maps next to the anomaly maps. We derive the binarization threshold by optimizing for the best possible dice score.}
\label{fig:recos_zoom}
\end{figure*}

\begin{figure*}[!h]
    \centering
    \includegraphics[width=\linewidth]{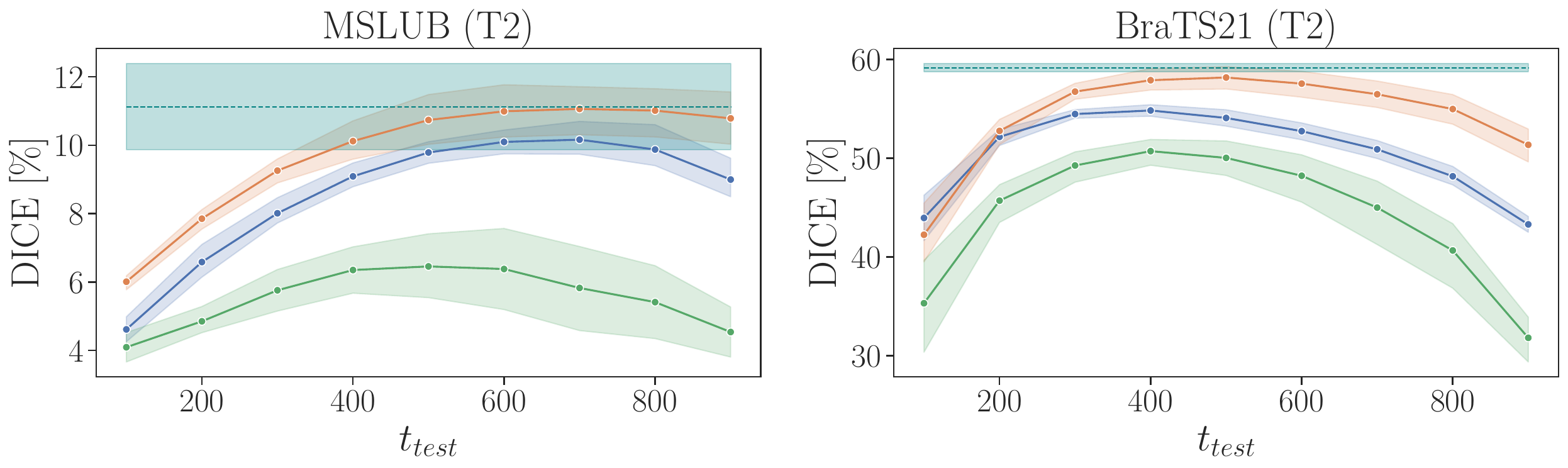}\vfill
    \includegraphics[width=\linewidth]{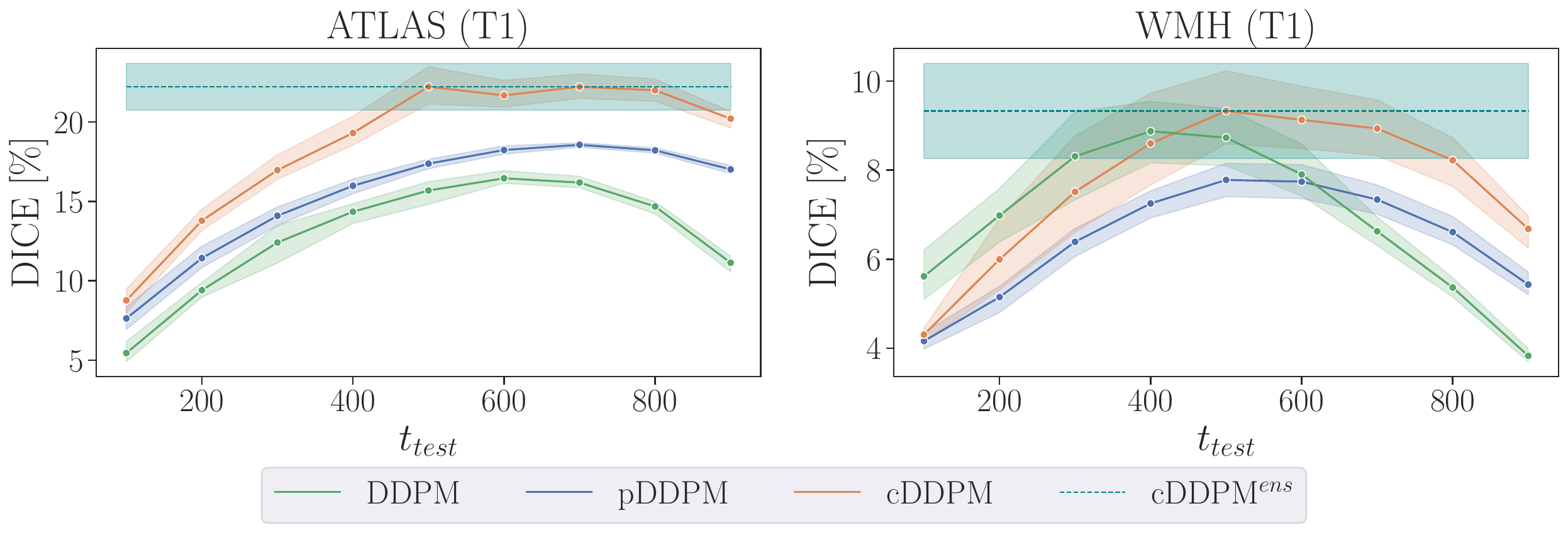}
    \caption{Comparison of different noise levels t$_{test}$ regarding the AUPRC. Top row: MSLUB (left) and  BraTS21 (right) data sets. Bottom row: ATLAS (left) and WMH data set. The superscript $ens$ denotes the ensembling of reconstructions from different noise levels $t_{test}$ in $[250, 500, 750]$. }
    \label{fig:ablations_noise}
\end{figure*}
We first compare the overall reconstruction quality of healthy and unhealthy structures. Second, we evaluate the domain adaptation properties of our approach and investigate the effect of our conditioning mechanism. Lastly, we evaluate the segmentation performance for different data sets, comparing our approach to established state-of-the-art UAD methods. 
\subsection{Reconstruction Quality}
In Table \ref{tab:recoresults} we compare baseline models regarding their ability to reconstruct the healthy anatomy, given the held-out test set of the IXI data set. Overall, \textit{DAEs}, \textit{pDDPMs} and cDDPMs show high performance regarding the image-based similarity metrics. In contrast, lower reconstruction quality is reported for the dense autoencoder-based baselines. Comparing the DDPM-based approaches, both \textit{pDDPM} and cDDPM outperform the baseline \textit{DDPM} in terms of reconstruction quality with statistical significance ($p<0.05$). We also analyze the unhealthy-to-healthy error ratio ($l1$-ratio) based on the unhealthy test sets. Generally, a higher $l1$-ratio is preferable as this indicates that the model successfully reconstructs the healthy anatomy without replicating the unhealthy parts of the input. Overall, the $l1$-ratio is highest for cDDPM across almost all data sets, except for the WMH, where \textit{DDPM} shows competitive performance. While the $l1$-ratio of \textit{DAEs} is high for the BraTS data set, it is substantially lower for all other data sets, indicating limited generalization. Additional results on the$l1$-error for all datasets are presented in Table \ref{tab:recohealthy} in the supplementary material.
\subsection{Conditioning Effect}
To evaluate the effect the additional conditioning input has on individual reconstructions, we simulate different conditioning inputs, varying in the amount of used image information. Furthermore, we apply artificial contrast levels for the input images to mimic domain shifts. For each conditioning input and contrast level, we provide the reconstructions generated by cDDPMs in  Figure \ref{fig:cond_eval} for a noise level of $t_{test}=500$ (50 $\%$). Moreover, we compare the reconstruction of DDPMs and cDDPMs in the extreme case of pure noise as input ($t_{test}=T=1000$ (100$\%$)). 
From Figure \ref{fig:cond_eval}, it becomes evident that the overall shape of the brain to reconstruct is preserved across the different (masked) conditioning images.
However, the local intensity information in the reconstruction is only captured at regions covered in the conditioning image. In the extreme case of a noise level of 100$\%$, the reconstructions of cDDPMs still coarsely follow the intensity distribution provided by the conditioning image, leading to a blurry reconstruction of the input image. In contrast, for unconditioned DDPMs, only a generic reconstruction that shares low similarity with the given input image can be obtained. Therefore, the conditioning mechanism primarily guides the generation process of cDDPMs to maintain local intensity distributions of the input image while detailed structural information is not captured.
\subsection{Domain Adaptation}
To evaluate the domain adaptation to different data sets, we consider the healthy IXI data set as an in-domain data set and the unhealthy BraTS21 data set as an out-of-domain one. Note that for the BraTS21 data set, we only consider regions annotated as healthy. Thereby, we ensure the evaluation of domain shifts regarding scanner and brain diversity, not domain shifts arising from unhealthy brain MRI structures. In Figure \ref{fig:histos_ixi_brats}, we provide the histograms of input and reconstructions of the healthy IXI data set (left) and the unhealthy BraTS21 data set (right). We observe that DDPMs show substantial discrepancies across input and reconstruction intensity distributions. Particularly for simulated contrast levels, the histograms deviate. In contrast, the intensity distribution of images reconstructed by cDDPMs exhibits higher similarities with the input intensity distribution for in- and out-of-domain data. Compared to DDPMs, cDDPMs decrease the KLD by a factor of 2.3, 4.0 and 17.0 across the contrast factors, respectively, considering the IXI data set.
\subsection{Segmentation Performance}
\begin{figure*}[!h]
    \centering
    \includegraphics[width=\linewidth]{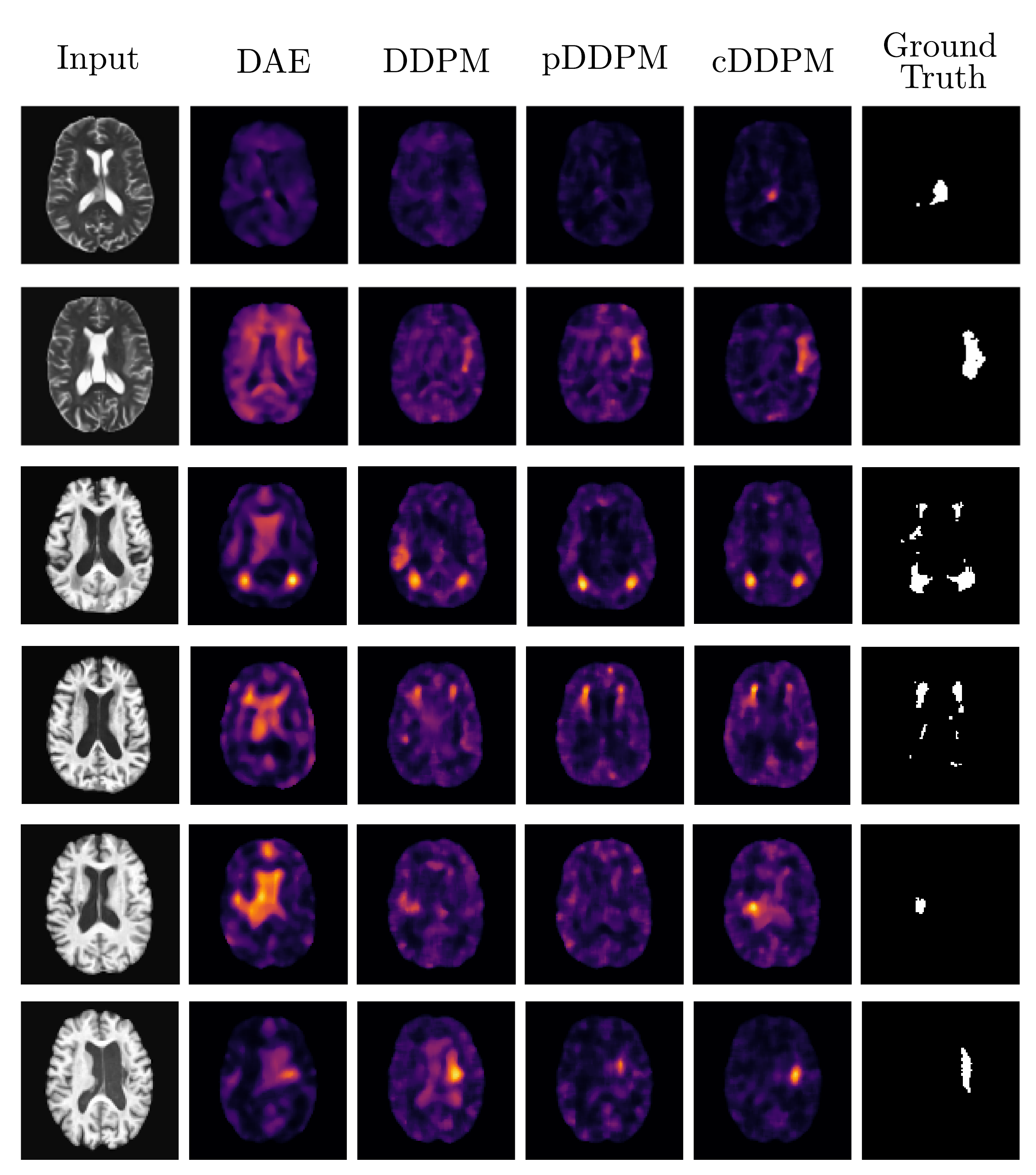}
    \caption{Exemplary residual maps from the BraTS21 data set (rows 1 and 2), the WMH data set (rows 3 and 4) and the ATLAS data set (rows 5 and 6). The input image, residual maps, and ground truth are shown from left to right. Corresponding segmentation maps are provided in Figure \ref{fig:pp_anomalymaps}.}
    
    \label{fig:app:recos}

\end{figure*}

We report the segmentation performance in Table \ref{tab:mainresults} considering all unhealthy test sets. For cDDPMs, improved performance is reported compared to all baselines across all data sets, except for the WMH data set, where the performance is on par with \textit{DDPMs}. 
While the improvements for the cDDPM are statistically significant for the BraTS21 and ATLAS data sets ($p<0.05$), no significant difference can be observed for the MSLUB and WMH data sets.  
Furthermore, we report enhanced performance of cDDPMs when pre-training the encoder (SSL checkmark in Table \ref{tab:mainresults}) and ensembling the reconstructions of different noise levels (ENS checkmark in Table \ref{tab:mainresults}) for most data sets. 
Notably, the inference time of cDDPMs is reduced by $\sim37\%$ compared to \textit{pDDPMs} and increased by $\sim2\%$ compared to \textit{DDPMs}.\\
The reconstruction-based \textit{AEs}, \textit{(S)VAEs}, \textit{RA} and \textit{PHANES} as well as the self-supervised \textit{PII} and the feature-based \textit{EDC} are outperformed by a margin when compared to DDPM-based approaches. While the feature-based methods \textit{RD} and \textit{FAE} and the reconstruction-based \textit{DAEs} show competitive performance considering individual data sets, limited generalization across all evaluated data sets is shown.
For a qualitative assessment of the conditioning mechanism, we provide reconstructions and the corresponding anomaly maps in Figure \ref{fig:recos_zoom}, comparing unconditioned \textit{DDPMs} to cDDPMs. cDDPMs provide accurate reconstructions of the target image with aligned intensity distributions across input and reconstructions. In contrast, the reconstruction of \textit{DDPMs} shows less details with a slight intensity shift. Hence, the anomaly map of cDDPMs shows a higher contrast across normal and abnormal regions, which facilitates the delineation of the present pathology.\\ 
In Figure \ref{fig:ablations_noise} we provide an ablation study considering different noise levels applied during the diffusion process at test time. The achieved AUPRC score is dependent on the noise level. Applying the ensembling strategy reduces this dependency and achieves consistent performance without specifying an individual noise level. \\ We supply a collection of exemplary residual maps for different baseline methods in Figure \ref{fig:app:recos}. Compared to the baselines, our proposed cDDPM demonstrates the low occurrences of false positives and a pronounced contrast in the residual maps. This observation is further supported by Figure \ref{fig:pp_anomalymaps} in the supplementary material.
\\ Additionally, we include ablation studies on the applied post-processing steps in the supplementary material.

\section{Discussion} \label{sec:discussion}
UAD in brain MRI has gained significant attention due to its potential to identify abnormalities without costly data annotation. Compared to supervised approaches that rely on annotated data sets, UAD methods take a different approach by learning the underlying data distribution of healthy brain anatomy and identifying anomalies as outliers. This is a crucial property for screening tasks, where any pathology has to be detected, even if it is not represented in an annotated training set. \\
In this study, we focus on reconstruction-based UAD with DDPMs. DDPMs generate images by reconstructing an input corrupted by noise, leveraging the high-dimensional latent space to achieve high-fidelity reconstructions of fine-grained structures. However, we show that the forward and backward processes of DDPMs do not sufficiently capture the highly variable intensity characteristics of MRI scans. This results in differences between input and reconstructions that are difficult to distinguish from differences that arise from actual pathologies. This limitation becomes especially prominent in the presence of domain shifts at test time.
\\
To address this challenge, we propose conditioned DDPMs (cDDPMs) for UAD in brain MRI. We train a DDPM to reconstruct healthy brain anatomy and condition the denoising process by a latent feature representation of the input image derived by an additional image encoder. While the additional dense feature representation can capture local intensity information of the image to reconstruct, it is unsuitable for the reconstruction of detailed structures \cite{baur2021autoencoders,bercea2023generalizing}. This is important, as providing detailed structural information of the unhealthy input image could lead to copying abnormal structures, which would prevent the detection of pathologies. As demonstrated in our results, our approach facilitates the effective utilization of the conditioning signal, contributing to improved reconstructions that adapt locally to the input intensity distribution. Therefore, we observe enhanced domain adaptation capabilities to both real and simulated intensity profiles with our conditioning mechanism. Moreover, our results indicate that the additional conditioning signal does not support the replication of unhealthy structures.  Finally, these individual features of our approach lead to an improved segmentation performance across various data sets.\\
In the following, we systematically evaluate our approach regarding reconstruction quality, domain adaptation, and segmentation performance based on five different data sets.  

\subsection{Reconstruction Quality} \label{subsec:reco}
We compare the reconstruction quality of our method with baseline models on the healthy IXI data set in Table \ref{tab:recoresults}. 
For the \textit{AE} and \textit{(S)VAE}, overall the worst reconstruction quality is reported. A reason for this is seen in the strict bottleneck enforced by the dense latent space as it inhibits information flow \cite{baur2021autoencoders}. In contrast, methods like \textit{DAE} or \textit{DDPMs} that are not constrained by a dense latent space but by a denoising task \cite{kascenas2021denoising} show improved reconstruction performance. While \textit{pDDPMs} and cDDPMs outperform the baseline \textit{DDPM} in terms of reconstruction quality, we observe that all models are outperformed by the \textit{DAE}.
We note that the overall training objective of the compared generative models is to reconstruct the image with high accuracy and copying the input image would be a trivial solution. However, for the UAD task, it is crucial that the given input image is not solely copied but that pseudo-healthy representatives replace unhealthy anatomy. Hence, comparing only the reconstruction quality of healthy anatomy does not necessarily reflect the usefulness of the UAD task. Therefore, we utilize the $l1$-ratio where high values indicate a better trade-off between the reconstruction of healthy and unhealthy anatomy and vice-versa. While \textit{DDPMs} and particularly the cDDPM achieve a high $l1$-ratio, across all unhealthy data sets, it becomes evident that the \textit{DAE} fails to generalize to different pathology types, a crucial property of UAD methods. A reason for that is seen in the chosen noise type in \textit{DAE} that mimics the visual appearance of tumors \cite{bercea2023generalizing,Lagogiannis.2023}. \\ In summary, the cDDPM shows improved reconstruction quality compared to \textit{DDPMs} while preserving a high $l1$-ratio. We conclude that the conditioning mechanism effectively captures intensity information from the input image for an improved reconstruction without providing too much detailed structural information that would enable the cDDPM to solely copy the input image. 

\subsection{Domain Adaptation}
We evaluate the domain adaptation capabilities of cDDPMs by simulating different contrast levels and conditioning inputs in Figure \ref{fig:cond_eval}. The reconstructions show that while the overall shape is preserved across different conditioning masks, meaningful reconstructions are achieved only in regions covered by the conditioning image. Particularly, the conditioning image is critical in capturing local intensity information, demonstrating the ability of cDDPMs to adapt to different contrast levels and effectively capture intensity information. This becomes even more evident when high noise levels are considered (100\%), where the only source of information concerning the given input image is the conditioning image. Here, the reconstruction becomes totally dependent on the shape and intensity characteristics of the conditioning image. The conditioning facilitates a blurred reconstruction of prominent local intensity changes. This indicates that the dense latent representation used for conditioning guides the reconstruction with local intensity details from the input image, while detailed structural information is not captured. Similar results are reported in the literature comparing AE architectures with dense and spatial latent spaces \cite{baur2021autoencoders}.
Furthermore, these findings indicate that cDDPMs effectively learn to balance information from the noisy input image and the conditioning encoder features during training, adapting according to the input's noise level.
\\
We explore the domain adaptation capabilities of cDDPMs in real-world scenarios where a different, out-of-domain data set is used for testing. To assess the domain adaptation ability, we investigate the deviation between the intensity distributions of the input and reconstruction by plotting histograms and calculating the Kullback-Leibler Divergence (KLD) as a proxy in Figure \ref{fig:histos_ixi_brats}.
Our results show that cDDPMs exhibit improved performance in capturing and estimating the intensity distribution. Particularly when simulating contrast levels considerably different from the training distribution, cDDPMs demonstrate superior alignment of the histograms and lower KLD values compared to unconditioned \textit{DDPMs}.
This analysis highlights the potential of the conditioning mechanism in cDDPMs to effectively adapt to unseen variations in intensity profiles and improve the coherence between input and reconstruction, which can contribute to improved domain adaptation. 

\subsection{Segmentation Performance}
Overall, cDDPMs demonstrate competitive or superior results compared to traditional autoencoder-based approaches, as well as the baseline \textit{DDPM} and \textit{pDDPM}, as presented in Table \ref{tab:mainresults}. Additionally, the feature-based and self-supervised methods are outperformed by our cDDPM. We conclude that the enhanced reconstruction quality, global and local intensity information capture, and effective domain adaptation capabilities attributed to our conditioning approach contribute to strong segmentation performance. Furthermore, we observe that pre-training the encoder $F_{enc}$ slightly enhances the segmentation performance in most cases, indicating that starting from an already learned representation space has the potential to improve the overall integration of the conditioning features, compared to simultaneously training the parameters of both, DDPM and $F_{enc}$ from scratch.
Our results demonstrate improved performance of cDDPMs over \textit{pDDPMs}, indicating that the proposed conditioning mechanism is more effective compared to the patching strategy in \textit{pDDPMs}. A reason for this is seen in potential artifacts, introduced by the patching strategy. Furthermore, cDDPMs reduce complexity and inference time as there is no need for a costly patching strategy, making them a practical and efficient solution for UAD in brain MRI. 
Despite the superior reconstruction quality of the \textit{DAE} on healthy data, cDDPMs show stronger segmentation results given the final UAD task. A reason for this is seen in the \textit{DAEs}' ability to reconstruct unhealthy anatomy, particularly for pathologies differing from the appearance of tumors, as discussed in Subsection \ref{subsec:reco}. Notably, the performance of \textit{AEs} and \textit{VAEs} in this work is lower compared to previous studies such as \cite{baur2021autoencoders}. This discrepancy can be attributed to differences in the MRI sequences used. Unlike \cite{baur2021autoencoders}, which uses FLAIR images, we rely on T2- and T1-weighted images. Previous research has shown that hyperintense lesions in FLAIR images can often be detected using simple thresholding \cite{meissen2022challenging}. This indicates that the high performance of \textit{AEs} and \textit{VAEs} in \cite{baur2021autoencoders} may be largely due to the hyperintensity of the lesions, rather than the model’s reconstruction ability.  \\
To further analyze the effect of the conditioning mechanism, we compare reconstructions and anomaly maps of \textit{DDPMs} and cDDPMs in Figure \ref{fig:recos_zoom}. In contrast to \textit{DDPMs}, cDDPMs' reconstructions follow the local intensity information of the respective input images. This is in line with the observations in Figure \ref{fig:histos_ixi_brats}, where a lower distribution shift is reported for cDDPMs. This leads to a reduction of false positives, facilitating the delineation of anomalies. 
\\
In Figure \ref{fig:ablations_noise}, we explore the impact of noise levels on the segmentation performance. We demonstrate that cDDPMs outperform the baseline models across different noise levels for most data sets. However, we also observe that the noise level is a crucial hyper-parameter. High noise levels tend to result in more blurry and generic reconstructions, whereas low noise levels enable sharper reconstructions, including unhealthy anatomy. Thus, selecting an appropriate noise level is essential to achieve reasonable performance. However, the optimal value for the applied noise depends on the evaluated data set, as shown in Fig \ref{fig:ablations_noise}. We assume that the main reason for this dependency is the different sizes of pathologies in the considered data sets, as also stated in \cite{bercea2023mask}. We apply different noise levels and average the resulting reconstructions to address this dependency. Thereby, we utilize complementary information of different reconstructions and effectively mitigate the dependency on the noise level, which enhances the model's generalization abilities.\\

\subsection{Limitations and Future Work} 
Despite showing promising results, our approach has limitations. Specifically, the segmentation performance falls below that of supervised algorithms. However, compared to supervised networks, UAD methods offer a crucial advantage in detecting any pathologies unseen during training. While additional efforts are required for the practical clinical application of UAD methods, our approach demonstrates superior performance to state-of-the-art UAD methods, adding a valuable contribution to the field of UAD in brain MRI. 
\\
Another avenue for future work is the incorporation of multi-scale image encodings into our conditioning mechanism. Our study does not utilize multi-scale analysis, which could be advantageous in capturing fine-grained details of the intensity patterns and contextual information at different resolutions. By carefully integrating multi-scale image encodings without allowing a copy of the conditioning image, we see the potential to enhance the performance of our cDDPMs in capturing both global and local features of the input images. \\
We conduct our studies using a downsampled resolution. To improve performance, especially for detecting smaller lesions, future work could explore and evaluate our proposed method at higher resolutions. A promising approach to balance the increased computational cost is the use of latent diffusion models \cite{rombach2022high}.
To further enhance the conditioning mechanism, an additional direction for future work is to explore the use of 3D image encoders. In earlier studies, we have shown that 3D information can improve the reconstruction quality for VAEs \cite{bengs2021three,behrendt2022capturing} and expect similar improvements for DDPMs. Currently, our approach operates on 2D slices of the MRI data, which may limit the preservation of 3D context and spatial relationships between slices. By utilizing a 3D encoder to condition the 2D DDPM, we can potentially capture and preserve 3D contextual information. 

\section{Conclusion} \label{sec:conclusion}
UAD in brain MRI presents a promising alternative to supervised models, especially considering clinical screening tasks. DDPMs have demonstrated their utility for UAD, largely due to their high reconstruction accuracy of fine-grained structures. However, accurately reconstructing the intensity characteristics of a given MRI scan remains a challenge, especially when facing domain shifts.
To address this, we propose cDDPMs and introduce a conditioning mechanism that incorporates an additional feature representation of the input image into the DDPMs’ denoising process. Our findings indicate that this conditioning mechanism effectively addresses challenges of DDPMs regarding capturing accurate intensity capture and domain adaptation. 
As a result, our approach outperformed state-of-the-art architectures for UAD in brain MRI on various publicly available data sets. Our work addresses challenges of applying DDPMs for UAD in brain MRI and has practical implications for detecting and segmenting pathologies in scenarios where domain shifts are likely. 
\section*{Declaration of Competing Interest}
The authors declare that they have no known competing financial interests or personal relationships that influenced the work reported in this paper.
\section*{Acknowledgements}
This work was partially funded by the "Zentrales Innovationsprogramm Mittelstand" [grant numbers KK5208101KS0 and ZF4026303TS9] and by the Free and Hanseatic City of Hamburg (Interdisciplinary Graduate School).
\section*{Author Contributions Statement}
\textbf{Finn Behrendt:} Conceptualization, Methodology, Software, Writing- Original draft preparation. \textbf{Debayan Bhattacharya, Robin Mieling, Lennart Maack} Conceptualization, Methodology, Writing- Reviewing and Editing. \textbf{Julia Krüger, Roland Opfer:} Conceptualization, Writing- Reviewing and Editing, Funding acquisition \textbf{Alexander Schlaefer:} Conceptualization, Writing- Reviewing and Editing, Funding acquisition, Supervision.
\section*{Ethics Statement}
This study was conducted using publicly available, fully anonymized datasets, and no new data were collected or experiments involving human subjects were performed. All procedures adhered to relevant laws, institutional guidelines, and ethical standards for research.
\bibliographystyle{model2-names.bst}
\bibliography{references}
\newpage

\section*{Supplementary Material} 

\subsection*{Post-Processing Analysis}

In Table \ref{tab:postproc}, we provide an analysis of the applied post-processing steps by excluding individual post-processing steps from the evaluation protocol. We show that while the median filter shows to have a large effect, the other post-processing techniques only show minor changes. Moreover, no post-processing strategy consistently works for all models or data sets, motivating further research and a systematic study about the effect of different post-processing steps for UAD in brain MRI.

\begin{table}[h!]
\caption{Post-processing analysis for all data sets regarding AUPRC. \checkmark indicates the presence and \xmark indicates the absense of Connected Component (CC), Medianfiltering (MF) or Brain Eroding (BE) in the evaluation phase, respectively. For all models, the mean $\pm$ standard deviation of the AUPRC are provided. Color-coded absolute differences concerning the respective baseline models are provided in the brackets.}
\resizebox{\linewidth}{!}{

\begin{tabular}{lcccccccccccc}
\toprule
Model & CC & MF & BE &                                    BraTS21 (T2) &                                      MSLUB (T2) & ATLAS (T1) & WMH (T1)\\
\midrule
  DAE & \checkmark & \checkmark & \checkmark & 49.38 $\pm$ 4.18 &    4.47 $\pm$ 0.69 &    13.37 $\pm$ 0.62 &  8.54 $\pm$ 1.02 \\
  DAE & \xmark &     \checkmark &     \checkmark &    49.38 $\pm$ 4.18 (0.00) &     4.47 $\pm$ 0.69 (0.00) &  8.53 $\pm$ 0.24 (\textcolor{red}{-4.84}) &   7.31 $\pm$ 0.91 (\textcolor{red}{-1.23}) \\
  DAE &     \checkmark & \xmark &     \checkmark &  39.69 $\pm$ 3.68 (\textcolor{red}{-9.69}) &   3.92 $\pm$ 0.35 (\textcolor{red}{-0.55}) &  9.21 $\pm$ 0.32 (\textcolor{red}{-4.16}) &   7.01 $\pm$ 0.68 (\textcolor{red}{-1.53}) \\
  DAE &     \checkmark &     \checkmark & \xmark &  48.79 $\pm$ 4.31 (\textcolor{red}{-0.59}) &   3.84 $\pm$ 0.55 (\textcolor{red}{-0.63}) &   8.77 $\pm$ 0.28 (\textcolor{red}{-4.60}) &    6.80 $\pm$ 0.81 (\textcolor{red}{-1.74}) \\
\hdashline
 DDPM & \checkmark & \checkmark & \checkmark &  50.61 $\pm$ 2.92 &   6.27 $\pm$ 1.58 &   17.77 $\pm$ 0.47  & 8.89 $\pm$ 0.89 \\
 DDPM & \xmark &     \checkmark &     \checkmark & 50.68 $\pm$ 2.81 (\textcolor{green}{0.07}) &   6.25 $\pm$ 1.51 (\textcolor{red}{-0.02}) &  14.65 $\pm$ 0.3 (\textcolor{red}{-3.12}) & 10.27 $\pm$ 0.96 (\textcolor{green}{1.38}) \\
 DDPM &     \checkmark & \xmark &     \checkmark & 35.93 $\pm$ 2.27 (\textcolor{red}{-14.68}) &    5.40 $\pm$ 0.95 (\textcolor{red}{-0.87}) & 12.02 $\pm$ 0.45 (\textcolor{red}{-5.75}) &    8.80 $\pm$ 0.71 (\textcolor{red}{-0.09}) \\
 DDPM &     \checkmark &     \checkmark & \xmark &  50.47 $\pm$ 3.07 (\textcolor{red}{-0.14}) &    5.60 $\pm$ 1.31 (\textcolor{red}{-0.67}) & 15.11 $\pm$ 0.32 (\textcolor{red}{-2.66}) &    9.89 $\pm$ 1.00 (\textcolor{green}{1.00}) \\
\hdashline
 pDDPM & \checkmark & \checkmark & \checkmark & 55.08 $\pm$ 0.54 &  10.02 $\pm$ 0.36 &    17.84 $\pm$ 0.10 &  7.52 $\pm$ 0.56 \\
pDDPM & \xmark &     \checkmark &     \checkmark &  55.07 $\pm$ 0.53 (\textcolor{red}{-0.01}) & 10.06 $\pm$ 0.37 (\textcolor{green}{0.04}) &  12.14 $\pm$ 0.21 (\textcolor{red}{-5.70}) &   7.33 $\pm$ 0.39 (\textcolor{red}{-0.19}) \\
pDDPM &     \checkmark & \xmark &     \checkmark & 34.99 $\pm$ 0.43 (\textcolor{red}{-20.09}) &   7.35 $\pm$ 0.28 (\textcolor{red}{-2.67}) & 13.35 $\pm$ 0.22 (\textcolor{red}{-4.49}) &    7.62 $\pm$ 0.80 (\textcolor{green}{0.10}) \\
pDDPM &     \checkmark &     \checkmark & \xmark &   54.10 $\pm$ 0.57 (\textcolor{red}{-0.98}) &   8.63 $\pm$ 0.48 (\textcolor{red}{-1.39}) & 13.71 $\pm$ 0.28 (\textcolor{red}{-4.13}) &    7.3 $\pm$ 0.95 (\textcolor{red}{-0.22}) \\
\hdashline
cDDPM & \checkmark & \checkmark & \checkmark & 58.82 $\pm$ 1.56 &  10.97 $\pm$ 1.17 &   22.22 $\pm$ 1.15 & 9.26 $\pm$ 1.07 \\
cDDPM & \xmark &     \checkmark &     \checkmark & 58.84 $\pm$ 1.57 (\textcolor{green}{0.02}) & 11.22 $\pm$ 1.31 (\textcolor{green}{0.25}) &  19.92 $\pm$ 1.45 (\textcolor{red}{-2.30}) &   9.86 $\pm$ 1.18 (\textcolor{green}{0.60}) \\
cDDPM &     \checkmark & \xmark &     \checkmark &   39.70 $\pm$ 1.40 (\textcolor{red}{-19.12}) &   8.31 $\pm$ 0.98 (\textcolor{red}{-2.66}) & 16.56 $\pm$ 1.11 (\textcolor{red}{-5.66}) &   8.34 $\pm$ 0.76 (\textcolor{red}{-0.92}) \\
cDDPM &     \checkmark &     \checkmark & \xmark &   58.52 $\pm$ 1.62 (\textcolor{red}{-0.30}) &  10.03 $\pm$ 1.34 (\textcolor{red}{-0.94}) & 20.41 $\pm$ 1.41 (\textcolor{red}{-1.81}) &   9.36 $\pm$ 1.15 (\textcolor{green}{0.10}) \\
\bottomrule
\end{tabular}}

\label{tab:postproc}

\end{table}

\subsection*{Reconstruction Analysis}
\begin{table}
\centering
\caption{Comparison of reconstruction quality of healthy structures across different models and datasets.  The $l1$-error is computed using ground truth annotations of the healthy regions. For all metrics, the mean $\pm$ standard deviation across the different folds are reported. The arrow $\downarrow$ indicates that lower values are favorable. DDPM-based models are evaluated by ensembling different values for $t_{test}=[250,500,750]$}
\resizebox{\linewidth}{!}{
\begin{tabular}{lcccccccc}
\toprule
\multirow{2}{*}{\textbf{Model}}&\multicolumn{6}{c}{$\bm{l1\textbf{-error}}$ (e-3) $\downarrow$}\\
 \cmidrule(l){2-7} 
  & IXI (T2) &  BraTS21 (T2) & MSLUB (T2)& IXI (T1)  & ATLAS (T1) & WMH (T1)  \\

\midrule

   VAE &  32.32 $\pm$ 0.64    &                         31.52 $\pm$ 0.65 &               36.78 $\pm$ 0.79 & 39.47 $\pm$ 0.60 & 49.12 $\pm$ 0.84 &             51.22 $\pm$ 0.88 \\
  SVAE &   29.08 $\pm$ 0.16   &                         28.33 $\pm$ 0.10 &               33.21 $\pm$ 0.17 & 39.65 $\pm$ 0.50 & 51.51 $\pm$ 0.87 &             54.46 $\pm$ 1.14 \\
    AE &   31.67 $\pm$ 0.41    &                         30.95 $\pm$ 0.40 &               35.77 $\pm$ 0.34 & 39.04 $\pm$ 0.38  & 48.18 $\pm$ 0.50 &              50.08 $\pm$ 0.66 \\
    RA &  34.36 $\pm$ 1.43    &                         34.09 $\pm$ 1.71 &               37.89 $\pm$ 1.52 & 40.66 $\pm$ 2.46  &48.25 $\pm$ 2.11 &             49.21 $\pm$ 1.88 \\
PHANES &   38.70 $\pm$ 1.74   &                        35.19 $\pm$ 1.75 &               41.64 $\pm$ 2.05 & 44.03 $\pm$ 1.32 & 55.14 $\pm$ 1.23 &             58.31 $\pm$ 1.33 \\
   DAE &  8.14 $\pm$ 0.17    &                        11.52 $\pm$ 0.28 &               10.74 $\pm$ 0.14 & 9.84 $\pm$ 0.45 & 15.72 $\pm$ 0.52 &             14.92 $\pm$ 0.46 \\
   \midrule
  DDPM &  14.29 $\pm$ 0.32    &                         18.01 $\pm$ 1.10 &               16.54 $\pm$ 0.54 & 16.03 $\pm$ 0.17 & 23.4 $\pm$ 1.78 &              23.78 $\pm$ 1.89 \\
 pDDPM &  9.70 $\pm$ 0.43    &                        12.14 $\pm$ 0.32 &               11.52 $\pm$ 0.52 & 11.21 $\pm$ 0.28 & 19.03 $\pm$ 0.23 &             19.16 $\pm$ 0.36 \\
 cDDPM &   9.68 $\pm$ 0.16   &                        12.26 $\pm$ 0.17 &               11.54 $\pm$ 0.22 & 11.36 $\pm$ 0.44 & 16.83 $\pm$ 0.43 &             16.86 $\pm$ 0.47 \\
\bottomrule
\end{tabular}}
\label{tab:recohealthy}
\end{table}

In Table \ref{tab:recohealthy}, we evaluate the reconstruction quality for healthy brain regions. Overall, when comparing T1-weighted and T2-weighted images, T1-weighted images exhibit a slightly higher reconstruction error. Additionally, the external test sets (BraTS21, MSLUB, ATLAS, WMH) show increased reconstruction errors compared to the IXI test set. However, the overall performance trends are consistent with those observed in Table \ref{tab:recoresults}.    

\subsection*{MRI Scanner Details}
\begin{table}[htbp]
\centering
\caption{MRI scanner details and age statistics for various datasets used in this study. The table includes scanner model, field strength, echo time (TE), repetition time (TR), and flip angle values for different MRI sequences (T1 and T2). Some datasets do not have fully specified scanner parameters (missing information is denoted by - ). * indicates that demographic information is only available for a subset of the data.}
\resizebox{\linewidth}{!}{
\begin{tabular}{lclcccc}
\toprule
\textbf{Dataset} & \textbf{Age (mean/std)} & \textbf{Model} & \textbf{Field Strength [T]} & \textbf{Echo Time [ms]} & \textbf{Repetition Time [ms]} & \textbf{Flip Angle [°]}\\ 
\midrule
\multirow{3}{*}{IXI (T1/T2)} & \multirow{3}{*}{49.4/16.7} & Philips Intera  & 1.5 & 9.6 / 5725.8 & 4.6 / 100.0 & 8.0 / 90.0   \\ 
    &           & Philips Gyroscan Intera & 3.0 & 9.8 / 8178.3 & 4.6 / 100.0 & 8.0 / 90.0    \\
    &           & GE  & 1.5 & -&-&- \\ 
    \midrule
BRATS (T2) & 61.2/12.0* & - & - & -&-&- \\ \midrule
MSLUB (T2) & 39.3/10.1 & Siemens Magnetom Trio MR & 3.0 & 6000.0 & 120.0 & 120.0 \\ \hline
\multirow{2}{*}{ATLAS (T1)} & \multirow{2}{*}{-} & - & 1.5 & -&-&- \\
      &               & - & 3.0 & -&-&-\\ \midrule
\multirow{5}{*}{WMH (T1)} &\multirow{5}{*}{-} & Philips Achieva & 1.5 & 7.9& 4.5&-\\
    &               & Philips Ingenuity & 3.0 &  9.9& 4.6  &-\\
    &               & Siemens TrioTim & 3.0 &  300& 1.9 & -\\
    &               & GE Signa HDxtT & 1.5 &  12.3& 5.2& -\\ 
    &               & GE Signa HDxt & 3.0 &  7.8& 3.0 &-\\ 
    \bottomrule
\end{tabular}}
\label{tab:app:scanner}
\end{table}
\newpage

\subsection*{Segmentation Maps}
\begin{figure}[h]
    \centering
    \includegraphics[width=.9\linewidth]{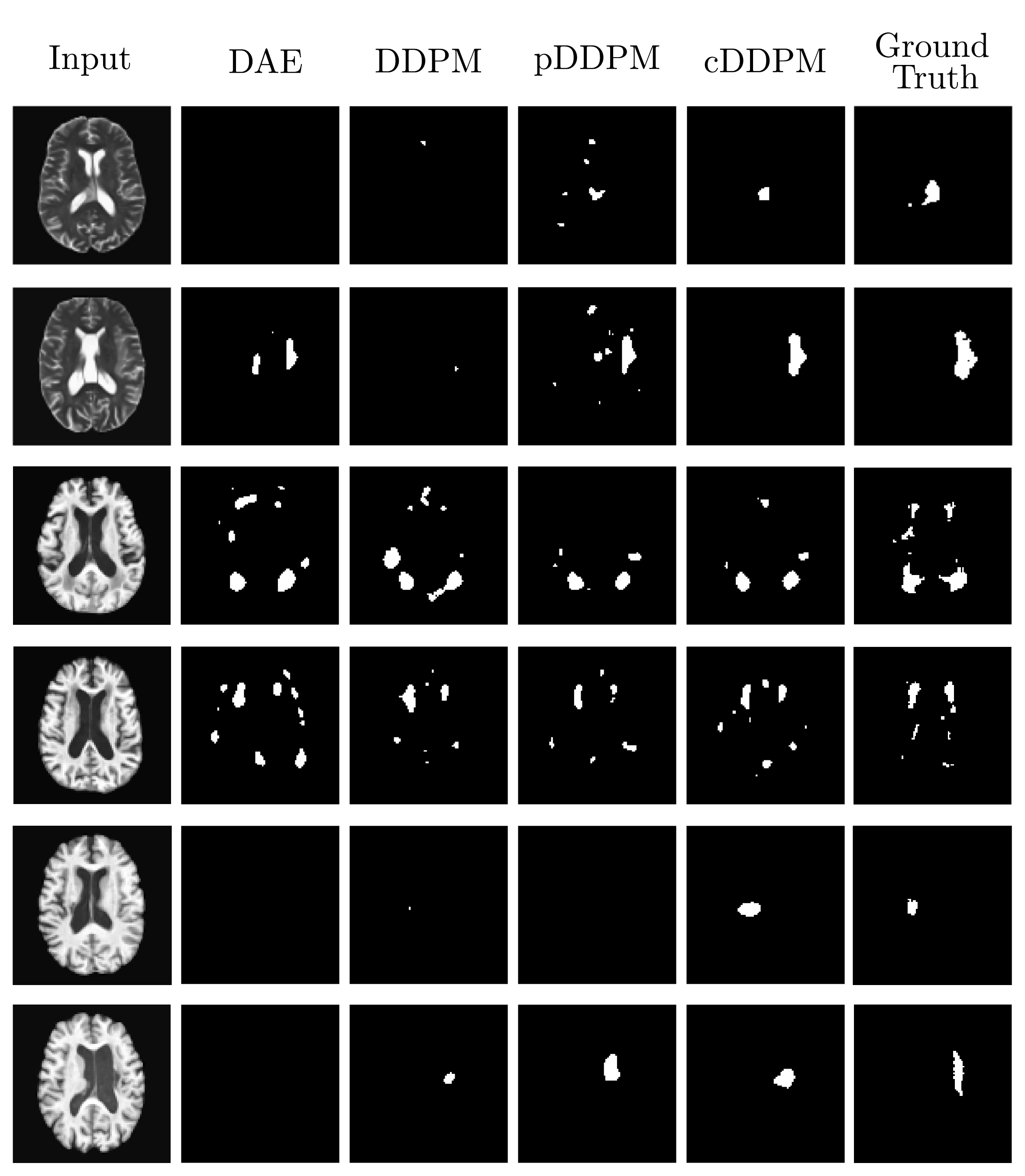}
    \caption{Segmentation maps corresponding to the residual maps in Figure \ref{fig:app:recos}, derived from the BraTS21 dataset (rows 1 and 2), the WMH dataset (rows 3 and 4), and the ATLAS dataset (rows 5 and 6). The binarization threshold for generating the segmentation maps was determined by optimizing for the highest possible Dice score.}
    \label{fig:pp_anomalymaps}
\end{figure}

\end{document}